\documentclass{jfm}

\usepackage{graphicx}
\usepackage{epsfig,psfrag}
\usepackage{natbib}
\usepackage{color}
\usepackage{mathrsfs}
\usepackage{amsbsy}
\usepackage{amssymb}
\usepackage[normalem]{ulem}
\usepackage{enumerate}

\ifCUPmtlplainloaded \else
\checkfont{eurm10}
\iffontfound
\IfFileExists{upmath.sty}
{\typeout{^^JFound AMS Euler Roman fonts on the system,
using the 'upmath' package.^^J}%
\usepackage{upmath}}
{\typeout{^^JFound AMS Euler Roman fonts on the system, but you
dont seem to have the}%
\typeout{'upmath' package installed. JFM.cls can take advantage
of these fonts,^^Jif you use 'upmath' package.^^J}%
}
\else
\fi
\fi
\ifCUPmtlplainloaded \else
\checkfont{msam10}
\iffontfound
\IfFileExists{amssymb.sty}
{\typeout{^^JFound AMS Symbol fonts on the system, using the
'amssymb' package.^^J}%
\usepackage{amssymb}%
 \let\leq=\leqslant
 \let\geq=\geqslant
}{}
\fi
\fi
\ifCUPmtlplainloaded \else
\IfFileExists{amsbsy.sty}
{\typeout{^^JFound the 'amsbsy' package on the system, using it.^^J}%
\usepackage{amsbsy}}
{\providecommand\boldsymbol[1]{\mbox{\boldmath $##1$}}}
\fi

\newcommand\bu{\mbox{\boldmath $u$}}
\newcommand\bx{\mbox{\boldmath $x$}}

\newcommand\beq{\begin{equation}}
\newcommand\eeq{\end{equation}}
\newcommand{\half}{\mbox{$\frac12$}}

\newcommand{\R}{{\cal R}}
\newcommand{\Sy}{{\cal S}}
\newcommand{\U}{{\cal U}}

\renewcommand{\vec}[1]{\mbox{\boldmath $#1$}}

\title[] {Minimal seeds for shear flow turbulence: using nonlinear transient
  growth to touch the  edge of chaos}

\author[C.C.T. Pringle, A.P. Willis and R.R. Kerswell]{
C\ls H\ls R\ls I\ls S\ns C.\ls T.\ns P\ls R\ls I\ls N\ls G\ls L\ls E$^1$,\ns 
A\ls S\ls H\ls L\ls E\ls Y\ns P.\ns W\ls I\ls L\ls L\ls I\ls S$^2$\break
\and \break
R\ls I\ls C\ls H\ns  R.\ns  K\ls E\ls R\ls S\ls W\ls E\ls L\ls L$^3$
}

\affiliation{$^1$Department of Earth Sciences, University of Bristol,
  Bristol, BS8 1RJ\\
$^2$School of Mathematics and Statistics, University of Sheffield,
Sheffield, S3 7RH\\
$^3$Department of Mathematics, University of Bristol, Bristol, BS8 1TW. }

\pubyear{???}
\volume{???}
\pagerange{???--???}
\date{\today}
\begin{document}

\maketitle

\begin{abstract}

We propose a general strategy for determining the minimal finite
amplitude disturbance to trigger transition to turbulence in shear
flows. This involves constructing a variational problem that searches
over all disturbances of fixed initial amplitude, which respect
the boundary conditions, incompressibility and the Navier--Stokes
equations, to maximise a chosen functional over an
asymptotically long time period.
The functional must be selected such that it
identifies turbulent velocity fields by taking significantly 
enhanced values compared to those for laminar fields. 
We illustrate this approach
using the ratio of the final to initial perturbation kinetic energies
(energy growth) as the functional and the energy norm to measure
amplitudes in the context of pipe flow. Our results indicate that the
variational problem yields a smooth converged solution providing the
amplitude is below the threshold amplitude for transition.  
This optimal is the nonlinear analogue of the well-studied (linear)
transient growth optimal. At and above this threshold, the optimising
search naturally seeks out disturbances that trigger turbulence by
the end of the period, and convergence is then practically impossible.
The first disturbance found to trigger turbulence as the amplitude is
increased identifies the `minimal seed' for the given geometry and
forcing (Reynolds number). We conjecture that it may be possible to
select a functional such that the converged optimal below threshold
smoothly converges to the minimal seed at threshold.  This seems at
least approximately true for our choice of energy growth functional
and the pipe flow geometry chosen here.

\end{abstract}

\section{Introduction\label{sec:intro}}

%
%
Shear flows are ubiquitous in our everyday lives yet predicting their
behaviour still remains an outstanding and important issue both
scientifically and economically. Typically such flows become turbulent
even though there may be an alternative linearly stable `basic
state', which is the simplest solution consistent with the driving
forces and boundary conditions. This bistability means that the
problem of transition comes down to understanding the
laminar-turbulent boundary in phase space that divides initial
conditions which lead to the turbulent state from those which relax back to the basic state.  This boundary has more generally been labelled
the `edge of chaos', allowing for transient turbulence 
(Skufca {\em et al.}\ 2006).  There have
been notable recent successes in tracking parts of this boundary
which, because it is a hypersurface in phase space, can be approached by
a simple bisection technique (Itano \& Toh 2001, Skufca, Yorke \&
Eckhardt 2006, Schneider, Eckhardt \& Yorke 2007, Duguet, Willis \&
Kerswell 2008).  As this technique is based upon integrating the
governing Navier-Stokes equations {\em forward} in time, only the parts of
the edge near (relative) attractors embedded in the edge are revealed
by this tracking approach.

Interestingly, these attracting regions invariably seem to be on
perturbation energy levels well above those known to be sufficient to
trigger transition. Viswanath \& Cvitanovic (2009) provide a good
illustration of this in a short pipe of length $\pi$ diameters, which
they numerically simulate with $85,715$ degrees of freedom. By only
mixing three fixed flow fields, they identify initial conditions which
experience an $O(10^4)$ magnification of the (perturbation) energy in
approaching a travelling wave thought to be embedded in the edge
relative attractor (Schneider, Eckhardt \& Yorke 2007, Pringle \&
Kerswell 2007): see their Table 4 for Reynolds number $Re=2000$.
Duguet, Brandt and Larsson (2010) tackle the same question in plane
Couette flow, finding evidence for energy growth {\em on} the edge of
over $10^2$ at $Re=400$ for a pair of oblique waves (see their Figure
9 where the plateau edge energy is $O(5 \times 10^{-3})$). Initial
conditions with low energies on the edge represent energy-efficient
targets to trigger transition, as an infinitesimal perturbation of
these states will lead to transition.  The most efficient of all 
perturbations will
be the flow field having the lowest energy $E_c$ on the edge,
hereafter called the {\em minimal seed} for transition, which
represents the closest (in perturbation energy norm) point of approach
of the edge to the basic state in phase space. This represents the
most dangerous disturbance to the basic state and as a result is of
fundamental interest either from the viewpoint of triggering
transition efficiently or, oppositely, in designing flow control
strategies.

Currently, there are no accepted strategies for identifying minimal
seeds beyond the impractical `brute force' approach of surveying {\em
  all} initial conditions.  The purpose of this paper is to continue
to develop a new strategy initiated in Pringle \& Kerswell (2010),
hereafter referred to as PK10, based upon identifying
finite-amplitude disturbance fields which, as they evolve via the {\em
  full} Navier-Stokes equations, 
maximise a key functional over a
period of time. This key functional is taken here to be the energy
growth of the disturbance over the time period as suggested in PK10
and Cherubini et al. (2010) in the boundary layer context. The
rationale behind this is the observation that the minimal seed must
experience considerable energy growth as it evolves in time up to the
attracting plateau on the edge. In the special case of a unique steady
relative attractor on the edge\footnote{For instance, in a pipe
  $\approx 2.5$ diameters long at $Re=2400$ and within the symmetric
  subspace $\R_2 \cap \Omega_2 \cap \Sy$ (see \S3.6, Duguet, Willis \&
  Kerswell 2008), the unique edge relative attractor is C3\_1.25
  (later renamed as N2 in Pringle, Duguet \& Kerswell 2009).}  the
minimal seed {\em will} be the optimal solution $\bu^*_0(\bx)$ to the
following variational problem: which initial condition {\em on the
  edge} (label this set $\Sigma$) will experience, for asymptotically
long times $T$, the largest energy growth defined as
\begin{equation}
{\cal G}(T):=\max_{\bu_0 \in \Sigma} \frac{\int \bu(\bx,T)^2
  \,dV}{\int \bu(\bx,0)^2 \, dV}
\label{impractical}
\end{equation}
where $\bu(\bx,t)$ is the flow at time $t$ evolved via the
Navier-Stokes equations from the initial condition
$\bu(\bx,0)=\bu_0(\bx)$ (incompressibility is tacitly assumed
throughout)
and $\bu=(u,v,w)$ is the
perturbation velocity field obtained by subtracting the laminar state
from the total velocity field (invariably used hereafter). The problem
with pursuing this criterion is not the fact that the relative edge
attractor may have a fluctuating energy as these fluctuations are
typically small compared to the total growth, but confining competitor
fields to the hard-to-define edge set $\Sigma$. A more practical
approach can be manufactured by turning the problem around to consider
the largest energy growth $G$ over all initial (incompressible)
conditions of a given perturbation energy $E_0$, that is
\begin{equation}
G(T;E_0):=\max_{\bu_0: \int \bu_0^2 \,dV=2E_0} \frac{\int \bu(\bx,T)^2
  \,dV}{\int \bu(\bx,0)^2 \,dV}.
\label{practical}
\end{equation}
%
At precisely $E_0=E_c$ where the edge touches the energy hypersurface at one velocity state, 
this optimisation problem considers the growth of this state (the minimal seed) against 
the energy growth of all the other initial conditions below the edge. 
Given that these latter initial
conditions lead to flows that grow initially but 
ultimately relax back to the basic state, 
it is reasonable to hypothesize that the
minimal seed remains the optimal initial condition for the revised
variational problem. 
A priori, the minimal seed energy $E_c(Re;{\rm
  geometry})$ is unknown but a very interesting quantity in its own
right, as its behaviour indicates how the basin of attraction of the
basic state shrinks with increasing $Re$. Hence, the variational
problem (\ref{practical}) must be solved as an increasing function of
$E_0$ until $E_c$ is reached. Knowing when this has occurred motivates
the following hypothesis.\\
\newline
\textbf{Hypothesis}: For asymptotically large $T$, the minimal seed
will be given by the flow field of initial energy $E_0$ which
experiences the largest energy growth such that for any initial
energies exceeding this $E_0=E_c$, the energy growth problem
(\ref{practical}) fails to have a smooth solution. \\
\newline
This is actually a very strong statement which really contains two
separate but related conjectures, the first being a necessary
condition for the second.\\
\newline
 \textbf{Conjecture 1}: For $T$ sufficiently large, the initial energy
 value $E_{fail}$ at which the energy growth problem (\ref{practical})
 first fails (as $E_0$ is increased) to have a smooth optimal solution
 will correspond exactly to $E_c$.\\
\newline
\textbf{Conjecture 2}: For $T$ sufficiently large, the optimal initial
condition for maximal energy growth at $E_0=E_c-\epsilon^2$ converges 
to the minimal seed at $E_c$ as $\epsilon \rightarrow 0$.\\
\newline 
The idea behind the first conjecture is that the optimisation
algorithm (\ref{practical}), in exploring the $E_0$-hypersurface for
the optimal solution, will detect any state on the energy hypersurface
that leads to turbulence, given that this leads to the highest values of
$G$. Once the algorithm is dealing with a turbulent endstate at time
$T$, the extreme sensitivity of the final state energy at $T$ to
changes in the initial condition, due to exponential divergence of
adjacent states, will effectively mean non-smoothness and prevent
convergence. Crucially, if true, this means that the failure of the
algorithm solving (\ref{practical}) should identify $E_c$ {\em
  regardless} of whether the minimal seed is the optimal solution of
(\ref{practical}) for $E_0=E_c$ or not. The key feature for Conjecture
1 to hold is not the precise form of the functional being maximised,
but the fact that the functional attains higher values for initial
conditions that go turbulent than those in the basin of attraction of
the basic state (other plausible choices include the final dissipation
rate, the total dissipation which has recently been explored with
success by Monokrousos et al (2011) in the context of plane Couette
flow, or more general Sobolev norms which emphasize strain
rates).  
The second, stronger conjecture, however, proposes that the
optimal initial condition converged at $E_0<E_c$ values, {\em
  approaches} the minimal seed as $E \rightarrow E_c$. This implies
that the energy growth functional is then a special choice which picks
out the minimal seed.  We present evidence in this paper to support
both these conjectures.

The variational problem (\ref{practical}) in the limit of
infinitesimally small energy $E_0$ reduces to the well-known (linear)
transient growth problem (Gustavsson 1991, Butler \& Farrell 1992,
Reddy \& Henningson 1993, Trefethen et al. 1993, Schmid \& Henningson
1994). In this case, the evolution of the
initial condition $\bu_0$ is determined by the linearised
Navier-Stokes equations and there are various ways to proceed:
e.g. for a matrix-based approach see Reddy \& Henningson (1993) and
for a time-stepping approach see, for example, Luchini (2000). The
optimal that emerges typically shows energy growth factors $G$ that
scale with $Re^2$ (when optimisation is also carried out over $T$) due
to the nonnormality of the linearised evolution operator. It has been
common practice to assume that this linear optimal (LOP) 
(for vanishing $E_0$) is a good
approximation to the minimal seed, as the energy of the minimal seed
is `small' compared to that of the basic flow or target turbulent
flow.  It was shown in PK10, however, 
that the presence of nonlinearity in the
variational problem is crucial in revealing new nonlinear optimals
(NLOPs) that emerge `in between' i.e.  for $0\, < \, E_0 \, <
\,E_c$. The calculations of PK10 were directed more at demonstrating the
feasibility of including nonlinearity in the transient growth
calculation and showing the dramatic manner in which this alters the
established linear result than identifying the minimal seed.  In
particular, $T=T_{lin}$ was taken where $T_{lin}$ is the optimal
growth time for the linear optimal (LOP) and is not asymptotically
large. Numerical limitations of the simulation code used in PK10
(written from scratch as part of the first author's thesis) also meant
that getting close to the edge proved difficult. In this paper, we
revisit those calculations using a well-tested parallel code
(described in Willis \& Kerswell 2009) to probe the `gap' between
$E_{fail}$ and $E_c$ noticed there.

The plan of the paper is as follows. Section \ref{sec:code} formulates
the variational problem (\ref{practical}) and describes briefly the
iterative scheme used to solve it. Section \ref{sec:mechanism}
analyses for the first time the mechanism by which the nonlinear
optimal (NLOP) which emerged in PK10 attains more growth than the
linear optimal (LOP). The results of PK10 are then extended to higher
energies in section \ref{sec:PK10} to re-examine the reported gap
between $E_{fail}$ and $E_c$. In section \ref{sec:longTime}, a larger
domain is studied using a longer optimisation time to provide a first
test of the conjectures discussed above. Our results are summarised
and discussed in section \ref{sec:conc} with a glossary of terms
following at the end.

\section{Formulation\label{sec:code}}

The context for our exploration is the problem of constant mass-flux
fluid flow through a cylindrical pipe. With length scales
nondimensionalised by half the pipe diameter $\half D$ and velocities
by the mean axial velocity $U$, the laminar flow is given by
\begin{equation}
\bu _{lam} = \U(s)\hat{\mathbf{z}} = 2(1-s^2)\hat{\mathbf{z}}
\end{equation}
using cylindrical coordinates $(s,\phi,z)$ aligned with the pipe axis.
In keeping with the majority of published work, results are reported
in time units of $D/U$. Energies are nondimensionalised by the
energy of the laminar flow in the same domain.  We then consider a
perturbation to this laminar profile such that the full velocity field
is given by
\begin{equation}
\U(s)\hat{\mathbf{z}} + \bu(s,\phi,z,t),
\end{equation}
where $\bu=(u,v,w)$ and  for convenience define the volume integral
\begin{equation}
\langle \ldots \rangle = \int_0^L \int_0^{2\pi} \int_0^1 \ldots 
 s \mathrm{d}s \, \mathrm{d}\phi \, \mathrm{d}z.
\end{equation}
In order to calculate the initial condition that produces the most
energy growth, we use a variational approach pioneered for the
linearised problem (Luchini \& Bottaro 1998, Andersson, Berggren \&
Henningson 1999, Luchini 2000, Corbett \& Bottaro 2000) but now
recently extended to incorporate the full Navier-Stokes equations
(PK10, Cherubini {\em et al.}\ 2010, 
Monokrousos {\em et al.}\ 2011: also see 
Zuccher {\em et al.}\ 2006 for earlier work using the boundary layer
equations). 
The functional we choose is defined as
\begin{eqnarray}
\mathscr{L}:=\langle \half \bu(\bx,T)^2\rangle &-&
\lambda \bigg[ \langle \half\bu(\bx,0)^2\rangle - E_0 \bigg] \nonumber \\
&-&\int_0^T \langle \boldsymbol{\nu} \cdot \bigg[
\frac{\partial \mathbf{u}}{\partial t} + \U\frac{\partial 
\mathbf{u}}{\partial z} + \U'u\mathbf{\hat{z}} 
- \bu \times \nabla \times \bu
+ \nabla p - \frac{1}{Re} \nabla^2 \mathbf{u} \bigg] \rangle \mathrm{d}t 
\nonumber \\ 
&-&\int_0^T \langle \Pi \nabla \cdot \mathbf{u} \rangle \mathrm{d}t 
-\int_0^T \Gamma(t) \langle \mathbf{u} \cdot \mathbf{\hat{z}} \rangle 
\mathrm{d}t.
\end{eqnarray}
This functional will be maximised by the same flow field as problem
\ref{practical}.  It is equivalent to finding the flow field with
greatest energy at time $t=T$, subject to four conditions applied
through Lagrange multipliers - namely that the initial condition,
$\bu(\bx,0)$ has kinetic energy $E_0$ and that it evolves subject to the
incompressible Navier-Stokes equations with fixed mass flux along the
pipe. The last constraint introduces a slight subtlety in that the
pressure field must be subdivided into a time-dependent constant
pressure gradient part $\Lambda (t) z$ which adjusts to maintain
constant mass flux and a strictly (spatially-) periodic part $\hat{p}$ so that
\begin{equation}
p:=\Lambda(t) z+\hat{p}(s,\phi,z,t).
\end{equation}
The Lagrange multipliers $\boldsymbol{\nu}=(\nu_s,\nu_\phi,\nu_z)$,
$\Pi$ and $\Gamma$ are known as the adjoint variables. The function
$\mathscr{L}$ will be maximised when all of its variational
derivatives are equal to zero. Taking variational derivatives leads us
to
\begin{eqnarray}
\delta \mathscr{L} = && \langle \, \delta \bu(\bx,T) 
\cdot [\bu(\bx,T)-\boldsymbol{\nu}(\bx,T) ]\,\rangle 
+ \langle \, \delta \bu(\bx,0) 
\cdot [ -\lambda\bu(\bx,0)+\boldsymbol{\nu}(\bx,0)]\, \rangle \nonumber\\
&-& \int_0^T \langle \delta \boldsymbol{\nu}
\cdot\bigg[\frac{\partial \mathbf{u}}{\partial t} + \U\frac{\partial 
\mathbf{u}}{\partial z} +\U'u\mathbf{\hat{z}} 
- \bu \times \nabla \times \bu + \nabla p 
- \frac{1}{Re} \nabla^2 \mathbf{u} \bigg] \rangle \mathrm{d}t
\nonumber \\
&+& \int_0^T \langle \delta \mathbf{u}
\cdot\bigg[\frac{\partial \boldsymbol{\nu}}{\partial t} + \U\frac{\partial 
\boldsymbol{\nu}}{\partial z} - \U'\nu_z\mathbf{\hat{s}} 
+ \nabla \times (\boldsymbol{\nu}\times\bu) - \boldsymbol{\nu} 
\times \nabla \times \bu + \nabla \Pi \nonumber \\
&& \hspace{8cm}+ \frac{1}{Re} \nabla^2 \boldsymbol{\nu} -\Gamma(t)
\mathbf{\hat{z}}\bigg] \rangle \mathrm{d}t \nonumber \\
&-& \int_0^T \langle \delta \Pi \nabla \cdot \mathbf{u} \rangle \mathrm{d}t 
- \int_0^T \delta \Gamma \langle \mathbf{u} \cdot \mathbf{\hat{z}} \rangle \mathrm{d}t
+ \int_0^T \langle \delta \hat{p}\nabla \cdot \boldsymbol{\nu} \rangle 
\mathrm{d}t
- \int_0^T \delta \Lambda (t) \langle \boldsymbol{\nu} \cdot
\mathbf{\hat{z}} \rangle \mathrm{d}t \nonumber\\
&-& \delta \lambda \bigg[ \langle \half\bu(\bx,0)^2\rangle - E_0 \bigg]. 
\label{eq:varL}
\end{eqnarray}
%
%
%
%
The nine terms making up the variational derivative can physically be
interpreted as meaning that to maximise $\mathscr{L}$: (i)
$\bu(\bx,T)$ and $\boldsymbol{\nu}(\bx,T)$ must satisfy a
compatibility condition; (ii) $\bu(\bx,0)$ and
$\boldsymbol{\nu}(\bx,0)$ must satisfy an optimality condition; 
(iii) $\bu$ must evolve according to the Navier-Stokes equations; 
(iv) $\boldsymbol{\nu}$ must evolve according to the adjoint Navier-Stokes equations;
(v)  $\bu$ is incompressible; (vi) $\bu$ has constant mass flux;
(vii) $\boldsymbol{\nu}$ is incompressible; (viii) $\boldsymbol{\nu}$
has constant mass flux; and (ix) the initial kinetic energy is $E_0$
(respectively as the terms appear in (\ref{eq:varL})\,).

In order to find a maximum to this problem, an iterative algorithm is
employed, seeded by an initial flow field $\bu_0:=\bu(\bx,0)$ of
appropriate kinetic energy (similar shorthand is used henceforth,
e.g. $\boldsymbol{\nu}_0:=\boldsymbol{\nu}(\bx,0)$). By integrating
this field forward in time in accordance with the Navier-Stokes
equations we can ensure that conditions (iii) and (v) are met. The
compatibility condition (ii) is satisfied by fixing
$\boldsymbol{\nu}_T=\bu_T$, which supplies a {\em final} condition for
the adjoint-Navier-Stokes equations to be integrated {\em backwards}
in time. This procedure generates $\boldsymbol{\nu}_0$ and ensures
conditions (iv) and (vi) are fulfilled. After this `forth-and-back'
time integration, the only outstanding Euler-Lagrange condition is
that the variational (Fr\'{e}chet) derivative
\begin{equation}
 \frac{\delta\mathscr{L}}{\delta\vec{u}_0} = -\lambda \vec{u}_0+
 \vec{\nu}_0
\end{equation}
should vanish.  As this does not happen automatically, $\bu_0$ is moved
in the ascent direction to increase $\mathscr{L}$ and hopefully
approach a maximum where it will vanish. An initial condition for the
next iteration is given by
\begin{equation}
\vec{u}_0^{n+1}=\vec{u}_0^n +\frac{\epsilon}{\lambda}\,
\frac{\delta\mathscr{L}}{\delta\vec{u}_0^n}   ,
\end{equation}
where $\langle(\delta\mathscr{L}/\delta\vec{u}_0)^2\rangle/\lambda^2
\sim \langle\vec{u}_0^2\rangle$ is a convenient rescaling.  The one
remaining Lagrange multiplier $\lambda$ is determined by arranging for
the new initial condition to satisfy the initial energy constraint
$\langle\frac{1}{2}(\vec{u}_0^{n+1})^2\rangle=E_0$. It is found that
$\lambda\sim\langle\vec{u}_T^2\rangle/\langle\vec{u}_0^2\rangle$, and
for the choice $\epsilon=1$ this strategy is equivalent to the power
method in the linear case (e.g. Luchini 2000).
In practice for the nonlinear case, the
choice $\epsilon=1$ is usually too large.  The following simple
strategy for the adaptive selection of $\epsilon$ was found to be
effective:
\begin{itemize}
\item Select an initial value for $\epsilon$, e.g.\ $0.5$.
\item Let 
\begin{equation}
d:=\biggl\langle
\frac{\delta\mathscr{L}}{\delta\vec{u}_0^n} \cdot
\frac{\delta\mathscr{L}}{\delta\vec{u}_0^{n+1}}\biggr\rangle \biggl/
\sqrt{ 
\biggl\langle
 \biggl( \frac{\delta\mathscr{L}}{\delta\vec{u}_0^n} \biggr)^2 
\biggr\rangle
\biggl \langle \biggl(\frac{\delta\mathscr{L}}{\delta\vec{u}_0^{n+1}}\biggr)^2
\biggr\rangle }
\biggr. ;
\end{equation}
\item if $d>0.95$ so successive adjustments in
  $\bu_0$ are essentially aligned then $\epsilon$ is doubled,
  otherwise, if $d<-0.5$ (anti-alignment) or
\begin{equation}
\biggl\langle\biggl(
\frac{\delta\mathscr{L}}{\delta\vec{u}_0^{n+1}} 
\biggr)^2 \biggr\rangle >4\,\biggl\langle \biggl(
\frac{\delta\mathscr{L}}{\delta\vec{u}_0^n} \biggr)^2 ,
\biggr\rangle
\end{equation}
whereby the derivative has becomes large, then $\epsilon$ is
halved.
\end{itemize}
Close to apparent convergence, when
$\langle(\delta\mathscr{L}/\delta\vec{u}_0^n)^2\rangle$ is very small,
a constant $\epsilon$ has sometimes been employed  to
  prevent multiple looping with tiny updates each time.


The numerical code used here is based on the well-tested code
described in Willis \& Kerswell (2009).  A Fourier decomposition is
employed in the periodic directions and a finite difference
approximation in the radial direction so that a typical dependent
variable is expanded as follows
\begin{equation}
A(s_n,\phi,z,t)=\sum_{m=-MM}^{MM} \, \, \sum_{l=-LL}^{LL} A_{nml}(t)
\exp(\mathrm{i}m \phi+\mathrm{i}\alpha l z) 
\qquad {\rm for} \quad  n=1,2,\ldots,NN 
\end{equation}
where $A$ is real so only half the coefficients ($m \geq 0$) need to
be stored, $\alpha=2\pi/L$ is the longest wavelength allowed by the
periodic axial boundary conditions, and $s_n$ are the roots of a
Chebyshev polynomial with finer resolution towards the wall. Typical
resolutions used were $(MM,NN,LL)=(23,64,11)$ for a $\half \pi\, D$
(PK10) pipe and $(23,64,37)$ for a $5\,D$ pipe. Using finite
differences in the radius is apt for parallelisation, which has been
implemented using MPI.  Time integration is performed using a second
order predictor-corrector method.

A fast (parallel) numerical code for handling the Navier-Stokes
equations and its adjoint is absolutely essential for successfully
implementing this iterative approach to optimisation. Each iteration
requires integrating the Navier-Stokes equations forward from $t=0$ to
$T_{opt}$ {\em and} the adjoint equations backwards from $t=T_{opt}$
to $0$, with typically $O(10^3)$ iterations required to be assured of
convergence.  There are also storage issues to circumvent, as the
adjoint Navier-Stokes equations, although linear in
$\boldsymbol{\nu}$, depend on $\boldsymbol{\bu}$. This either needs to
be stored in totality (over the whole volume and time period), which
is only practical for low resolution short integrations, or must be
recalculated piecemeal during the backward integration stage. This
latter `check-pointing' approach requires that $\boldsymbol{\bu}$ is
stored at regular intermediate points, e.g.\
$t=T_i:=iT_{opt}/n$ for $i=1,\ldots,n-1$, during the forward
integration stage. Then to integrate the adjoint equation backward
over the time interval $[T_{i},T_{i+1}]$, $\boldsymbol{\bu}$ is
regenerated starting from the stored value at $t=T_i$ by integrating
the Navier-Stokes equations forward to $T_{i+1}$ again. The extent of
the check pointing is chosen such that the storage requirement for
each subinterval is manageable. The extra overhead of this technique
is to redo the forward integration for every backward integration, so
approximately a $50\%$ increase in cpu time, assuming forward and
backward integrations take the same time.  As memory restrictions
may make full storage impossible, this is a small price to pay.

\section{The Nonlinear Energy Growth Mechanism \label{sec:mechanism}}

The basic ingredient for the strategy being explored here is the
solution of the variational problem (\ref{practical}) at a given
initial perturbation energy $E_0$. PK10 demonstrated the feasibility
of this and discovered that the nonlinearly-adjusted linear optimal
(LOP: see Schmid \& Henningson 1994) is quickly outgrown by a
completely new type of optimal (the NLOP) as $E_0$ increases from 0
(e.g.\ see figures 1 and 5 in PK10).  This NLOP exhibits both 
radial and azimuthal
localisation in a short $\half \pi D$-long pipe and
would undoubtedly also localise in the axial direction if the geometry
allowed. Localisation allows the perturbation to still retain
velocities of sufficient amplitude in an adequate volume that
nonlinearity is important while permitting the {\em global} energy to
be reduced. Without nonlinearity in the variational problem, any
localised state could be decomposed into {\em global} linear optimals
(e.g.\ by Fourier analysis), which would then evolve independently and,
all except the LOP, sub-optimally. PK10 remarked that the NLOP had a
two-phase evolution (e.g. see their figure 1) in which the
initially-3D optimal firstly delocalises (slices $a$ and $b$ in their
figure 2) followed by a second growth phase in which the flow becomes
increasingly 2D (streamwise-independent). We now examine this
evolution in more detail in order to understand how the NLOP is able
to achieve more growth than the LOP.

%
%
\begin{figure}                                                                
 \begin{center} \setlength{\unitlength}{1cm} 
 \resizebox{0.9\textwidth}{!}{\includegraphics[angle=0]{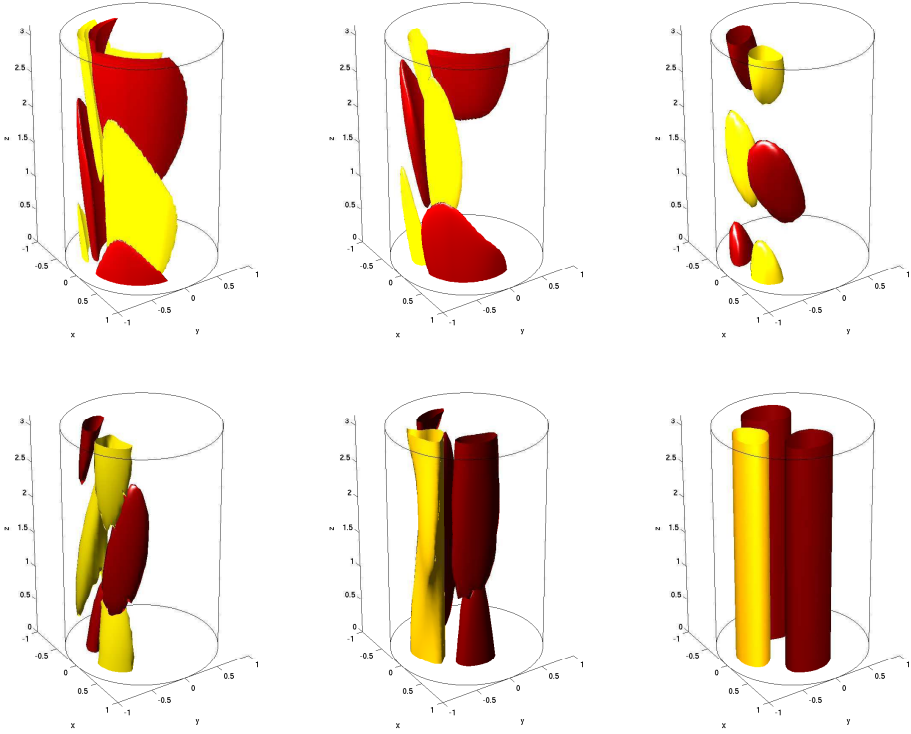}}
 \end{center}                                                   
\caption{The NLOP at $Re=1750$ and $E_0=1.8\times 10^{-5}$ calculated
  using resolution (MM,NN,LL)=$(23,64,11)$ at $t=0$ (upper left),
  $t=0.4$ (upper middle), $t=1$ (upper right), $t=2.5$ (lower left),
  $t=5$ (lower middle) and $t=10\,D/U$ (lower right). Isocontours are
  shown of the streamwise perturbation velocity: yellow (light) 50\%
  of the maximum and red (dark) 50\% of the minimum. Mean flow in each pipe 
  section is from bottom to top. }
\label{NLOPc}                                                                 
\end{figure} 
%
%
\begin{figure}
\setlength{\unitlength}{1cm} 
\centering \resizebox{\textwidth}{!}{\includegraphics{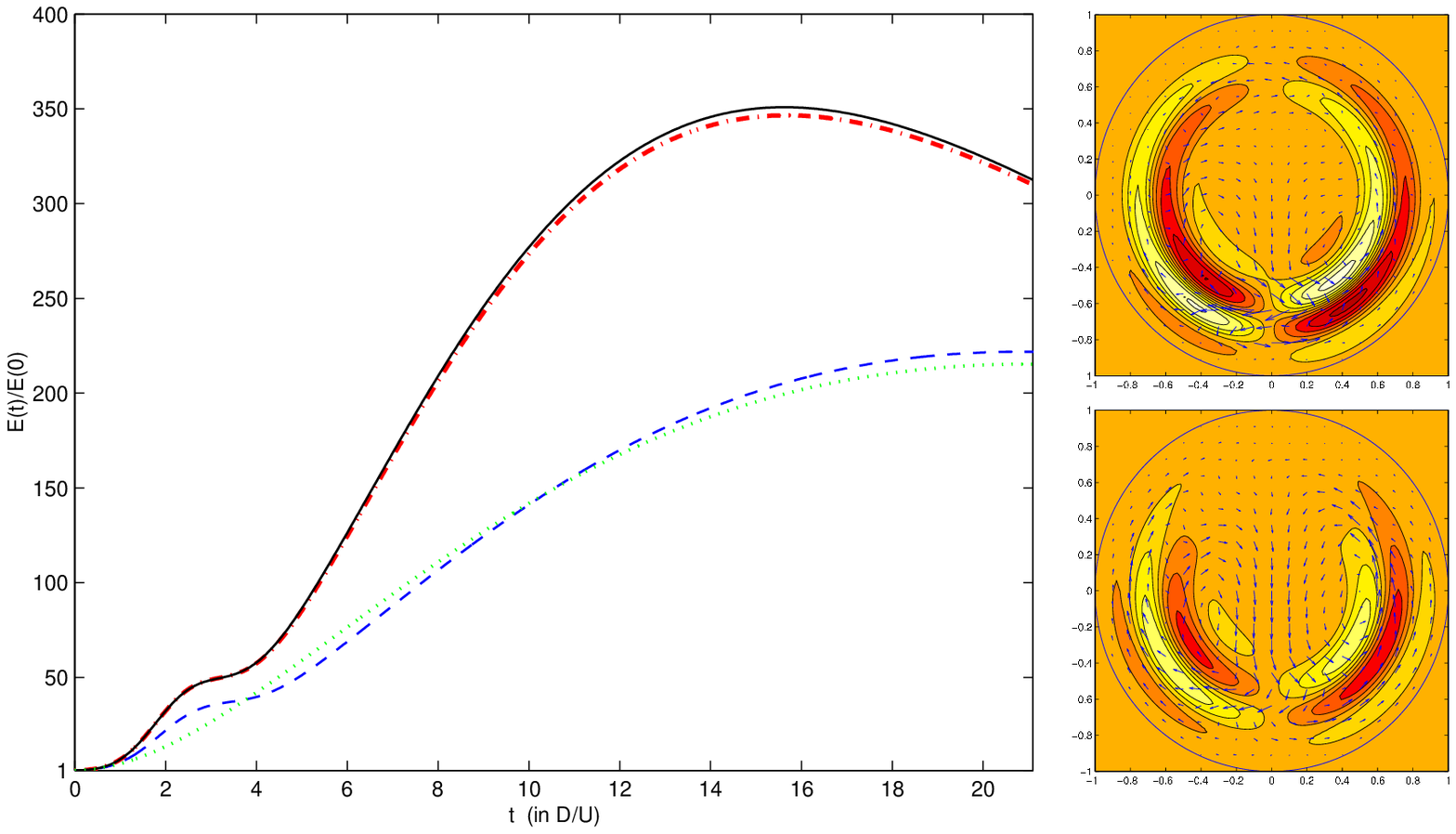}}
\caption{Left: energy growth $E(t)/E(0)$ against time $t$ for the NLOP
  of PK10 at Re=1750 and $E_0=1.8 \times 10^{-5}$
using full resolution (MM,NN,LL)=$(23,64,11)$ (black solid line) and
reduced resolutions $(7,64,1)$ (red dot-dash line) and $(2,64,1)$
(blue dashed line). Removing the helical mode effect by using
resolution $(1,64,1)$ (green dotted line) destroys the NLOP to leave
the LOP (notice the absence of a `shoulder' in the curve at $T \approx
3 \, D/U$). Right: NLOPs for $(7,64,1)$ (upper) and $(2,64,1)$
(lower). Contours indicate streamwise velocity perturbation (total
velocity with the laminar state of equivalent mass flux subtracted
off) using the same levels in both plots. Arrows indicate
cross-sectional velocities (same scale used for both). Note that the
slices are taken at the same point in the pipe but there may be a
phase difference between the solutions as they are calculated using
two different calculations. Also compare the slice for $(7,64,1)$ with
that for $(23,64,11)$ given in the upper left of figure \ref{NLOPa}.
}
\label{resgrowth}
\end{figure}


%
%
\begin{figure}                                                
\centering \resizebox{\textwidth}{!}{\includegraphics{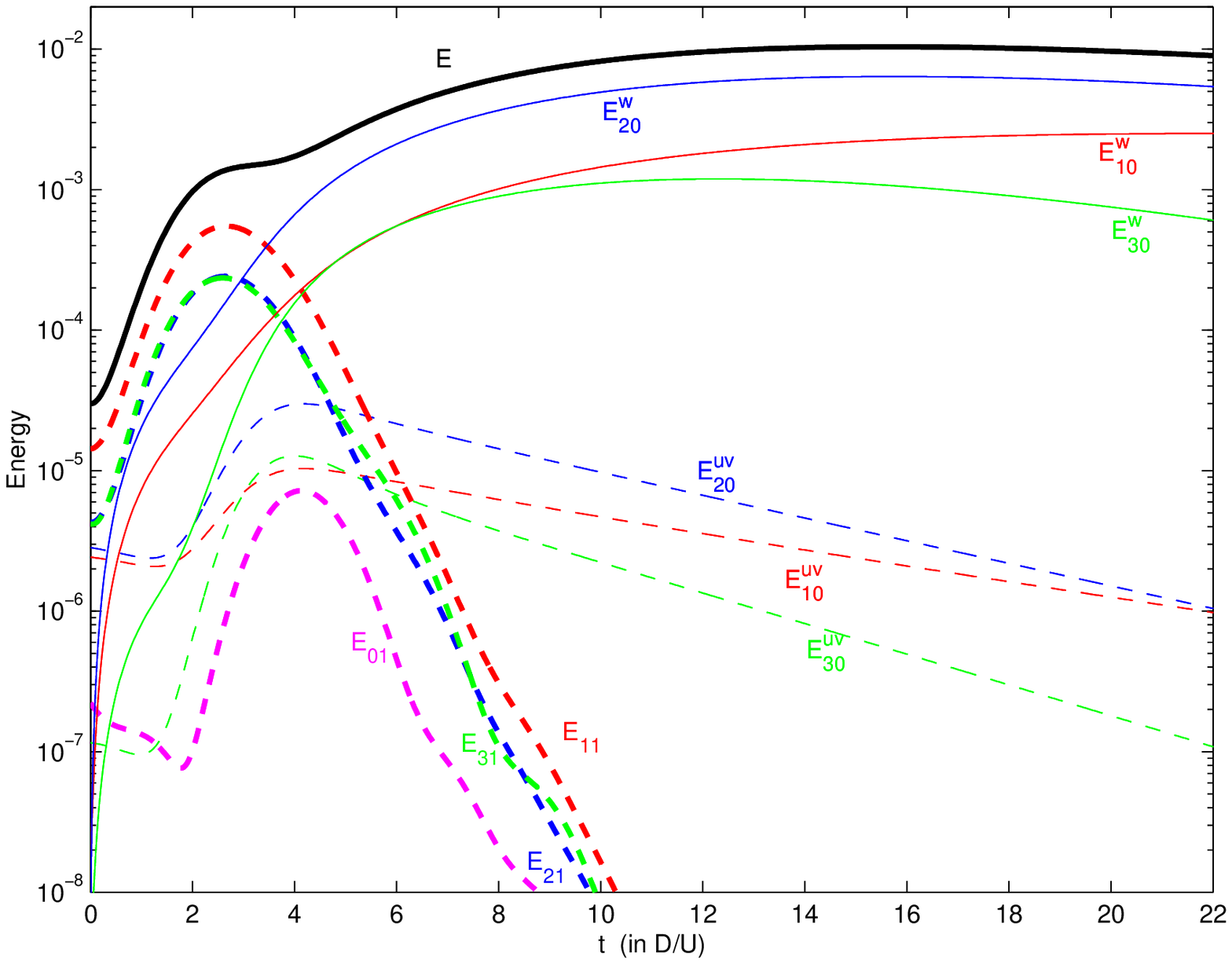}}
\caption{Modal energies for the NLOP run with resolution (7,64,1). $E$
  is the total perturbation energy; $E_{ml}^w$ indicates the energy in
  the $w$ (streamwise) velocity component with azimuthal and axial
  wavenumbers $m$ and $l$ respectively, and $E_{ml}^{uv}$ the energy
  in the $u$ \& $v$ (cross-stream) velocity components with azimuthal
  and axial wavenumbers $m$ and $l$ respectively. So $E_{m0}^w$ is the
  (streamwise) streak energy, $E^{uv}_{m0}$ is the (streamwise) roll
  energy and $E_{m1}$ is a  helical mode energy.}
\label{m7l1}                                                       
\end{figure}      

A first inspection of the 3D structure of the evolving NLOP actually
reveals 3 distinct phases of development.  Figure \ref{NLOPc} shows
how the axial structure of the NLOP evolves in time by plotting
isocontours of streamwise perturbation velocity along the pipe
(isocontours of streamwise vorticity show the same qualitative
behaviour). Initially the streaks are tightly layered and backward
facing i.e.\ inclined {\em into} the shear. By $t=0.4$, these layers
have been tilted into the mean shear direction (i.e.\ away from the
wall) by the shear and unpacked or separated slightly. This is the
inviscid Orr mechanism (Orr 1907) and gives an initial spurt of energy
growth. By $t=1$ the flow is then dominated by helical waves growing -
the `oblique' phase - before the flow becomes essentially although not
completely 2D by $t=10$ during the `lift-up' phase. This evolution
consisting of the Orr, oblique and lift-up phases in sequence is also
apparently seen for the critical disturbance found by Monokrousos et
al (2011) (D.S. Henningson, private communication).

\begin{figure}
 \begin{center} \setlength{\unitlength}{1cm} \begin{picture}(15,9.6)
 \put(0  ,4.8){\epsfig{figure=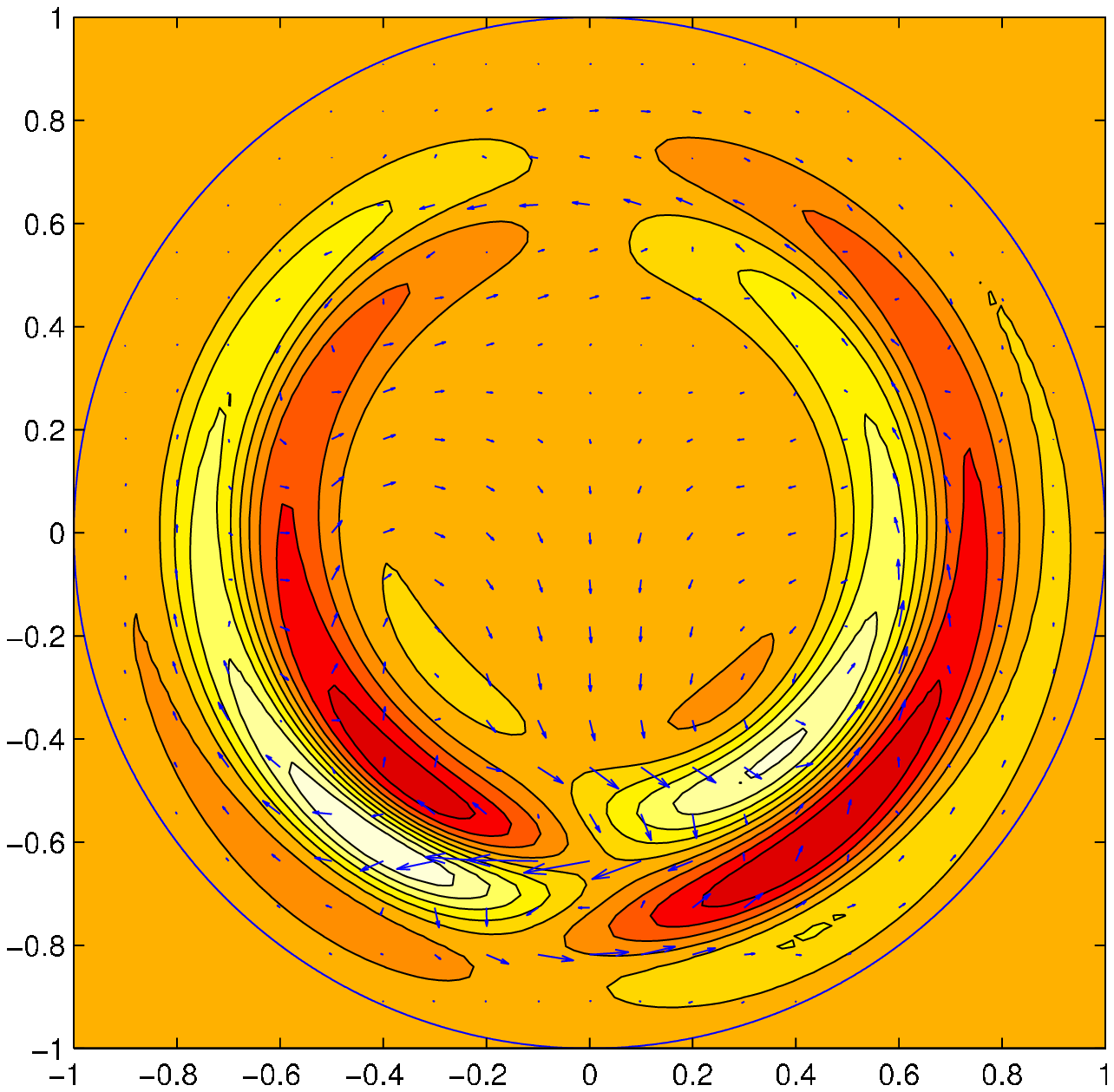,width=4.4cm,height=4.4cm,clip=true}}
 \put(4.6,4.8){\epsfig{figure=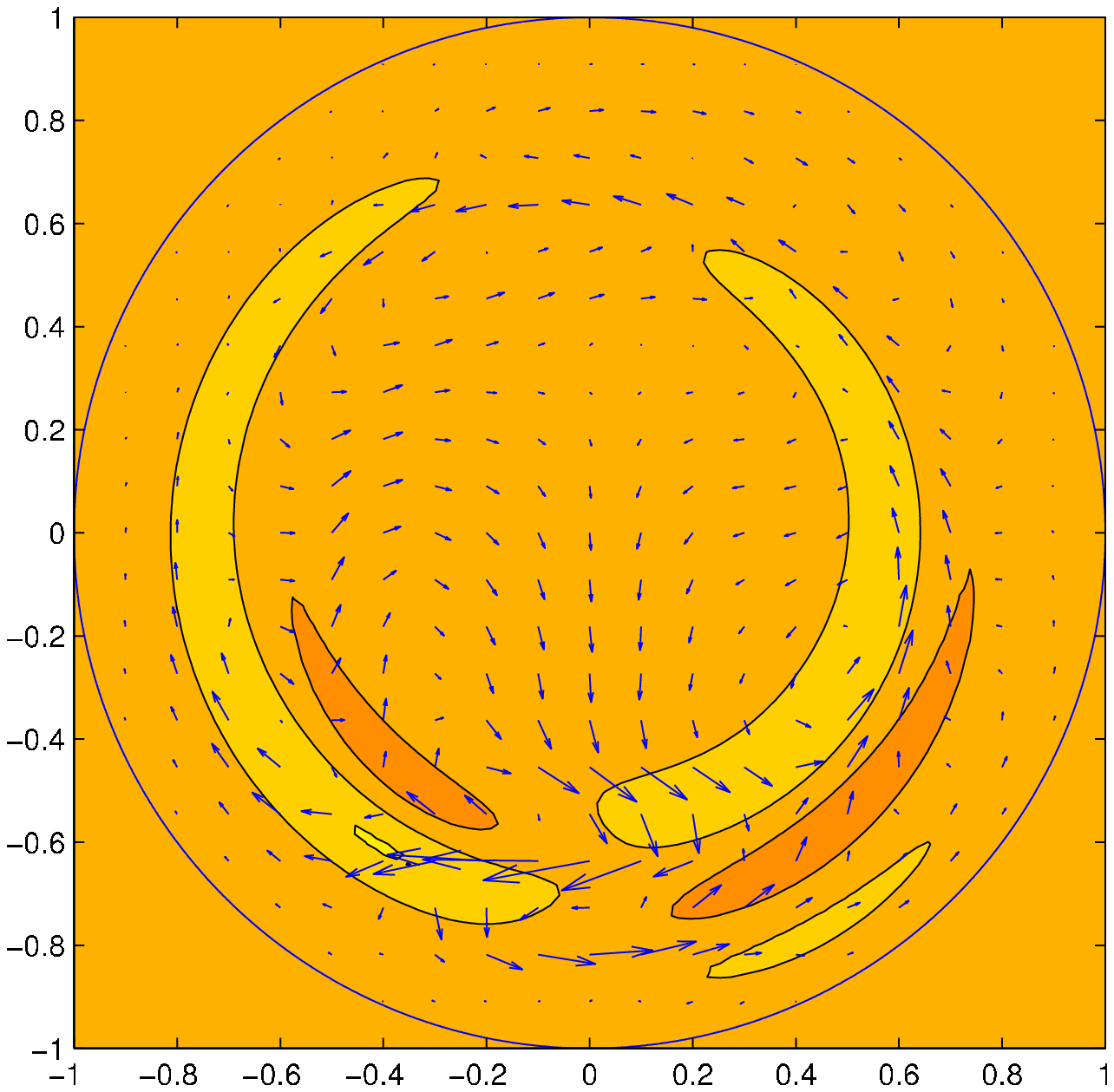,width=4.4cm,height=4.4cm,clip=true}}
 \put(9.2,4.8){\epsfig{figure=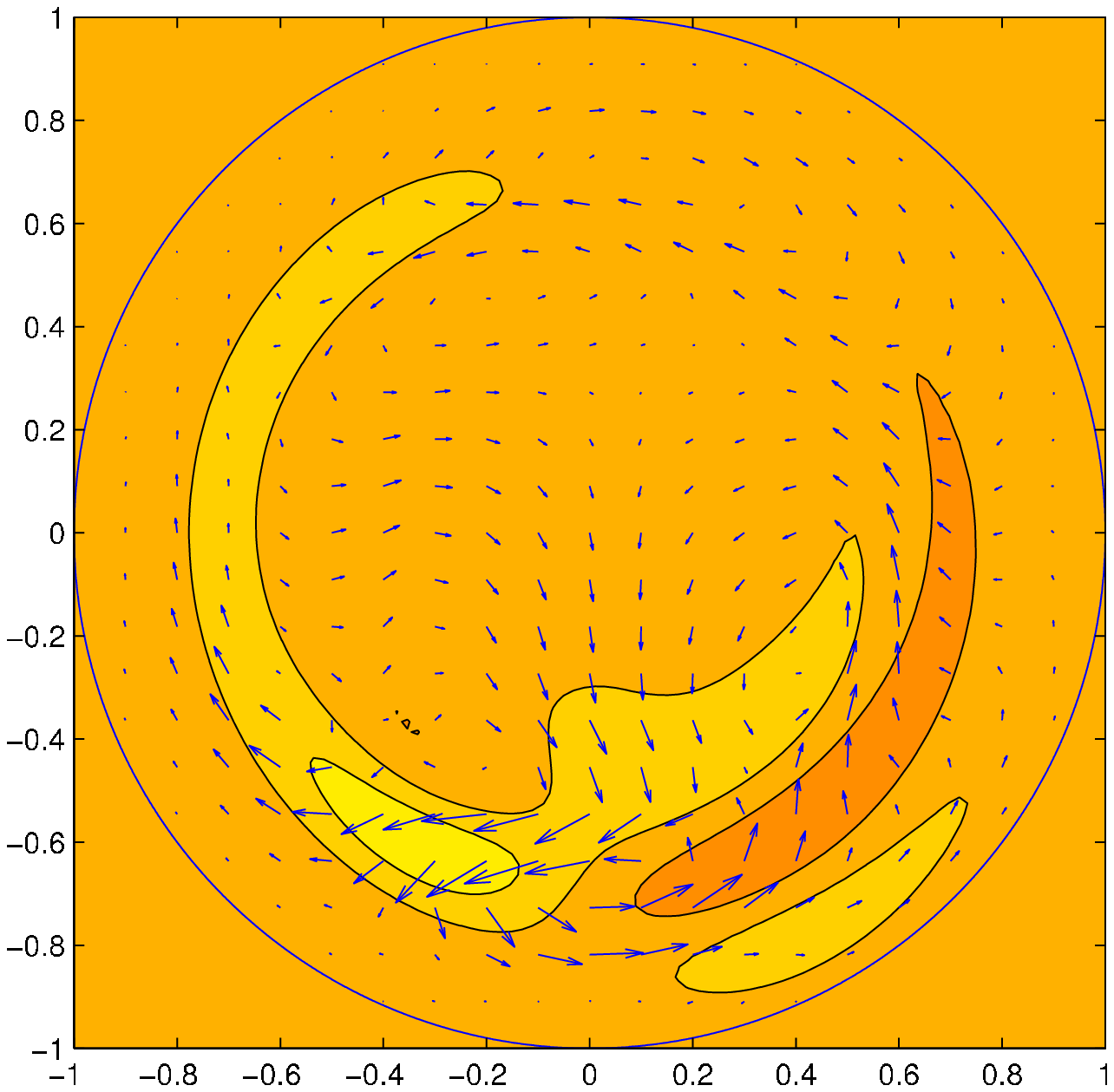,width=4.4cm,height=4.4cm,clip=true}}
%
 \put(0,0.2){\epsfig{figure=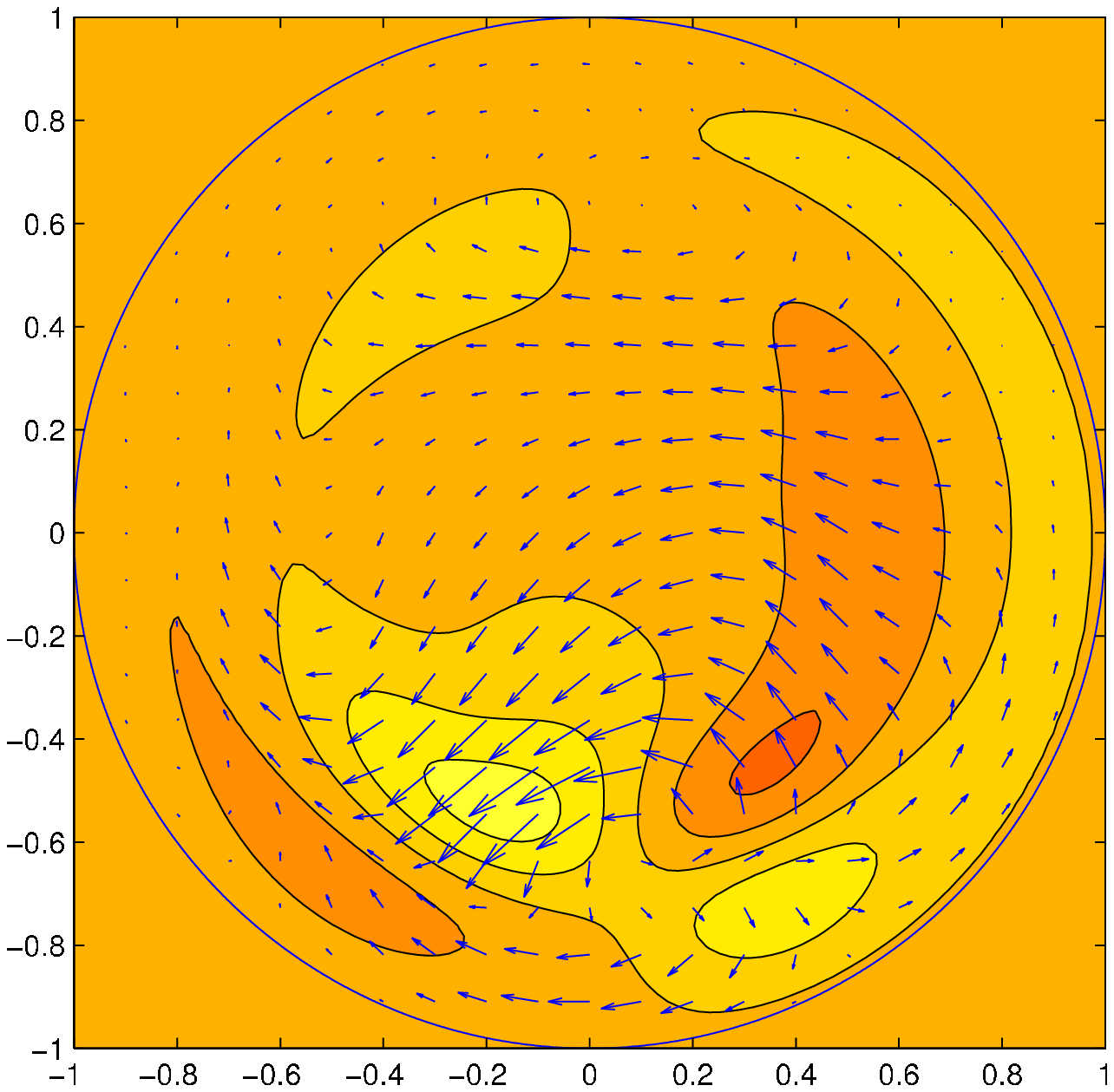,width=4.4cm,height=4.4cm,clip=true}}
 \put(4.6,0.2){\epsfig{figure=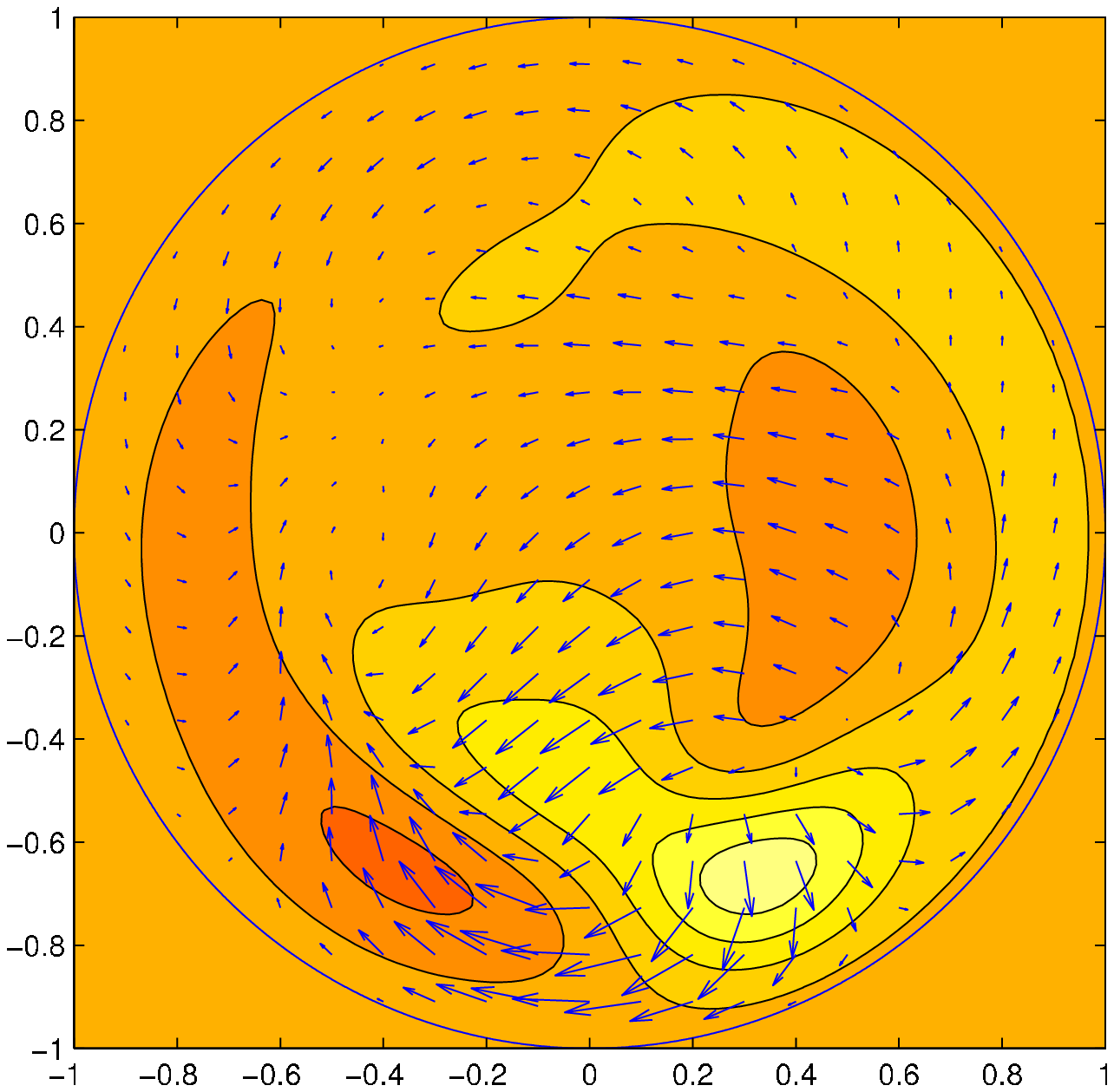,width=4.4cm,height=4.4cm,clip=true}}
 \put(9.2,0.2){\epsfig{figure=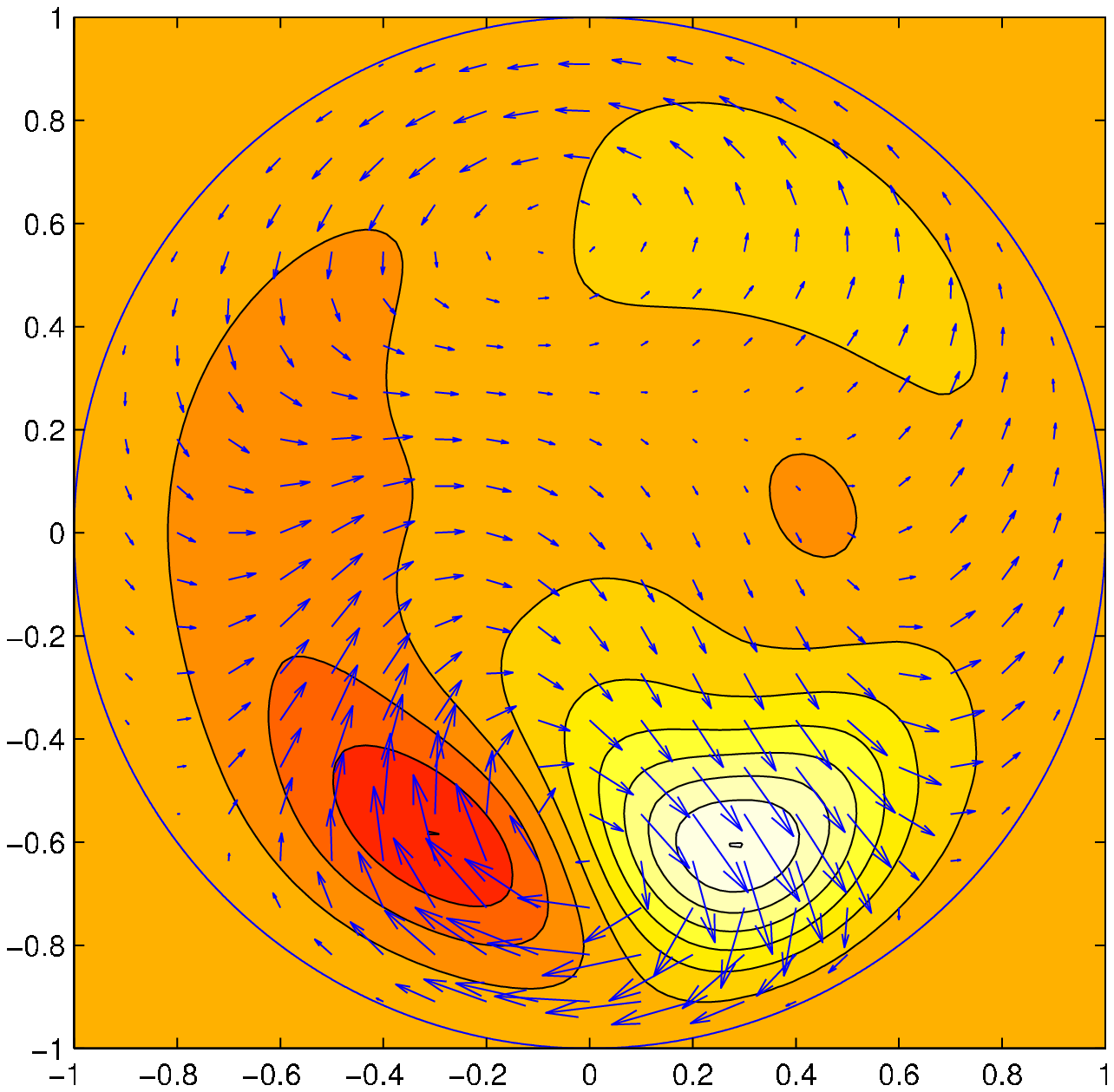,width=4.4cm,height=4.4cm,clip=true}}
%
 \end{picture} \end{center}
\caption{The NLOP at $Re=1750$ and $E_0=1.8\times 10^{-5}$ calculated
  using resolution (MM,NN,LL)=$(23,64,11)$ at $t=0$ (top left and top
  middle so same velocity field shown but with different contour levels),
  $t=0.2$ (top right), $t=0.6$ (bottom left), $t=0.8$ (bottom middle)
  and $t=1\,D/U$ (bottom right). Contours indicate the streamwise
  velocity perturbation (total velocity with the laminar state of
  equivalent mass flux subtracted off) and the arrows indicate the
  cross-stream velocity at a fixed slice in the pipe. All plots except
  the top left have 10 contour levels between the extremes of the
  streamwise velocity perturbation at $t=1$ and cross-stream
  velocities similarly scaled. The top left plot uses 10 contours
  between the extremes of the streamwise velocity perturbation {\em
    at} $t=0$ with arrows automatically scaled.}
\label{NLOPa}
\end{figure}

To clarify the oblique and lift-up phases, we reduce the considerable
degrees of freedom of the fully-resolved NLOP evolution down to those
that really matter.  As the Fourier-Fourier basis functions
$\exp(im\phi+i\alpha lz)$ in the velocity representation naturally
partition the linearized problem, we considered the optimal growth
calculation at $Re=1750$, $T_{opt}=T_{lin} \approx 0.0122Re=21.3\,D/U$
in a pipe length of $\half \pi D$ (so PK10 settings) at $E_0=1.8
\times 10^{-5}$ using the full resolution (MM,LL)=$(23,11)$ and
reduced resolutions $(7,1)$, $(2,1)$ and $(1,1)$ (full NN=$64$ radial
resolution was used for all). Figure \ref{resgrowth} shows how the
growth evolves as a function of time for each optimal initial
condition. All the calculations bar that for $(1,1)$ show the
distinctive `shoulder' in the growth centred at $t \approx 3 \, D/U$
(the time units $D/U$ will be suppressed hereafter), which signifies
the end of the (what PK10 called the `first') delocalisation phase and
the start of the next phase. The $(1,1)$ calculation fails to capture
the NLOP at all so that the optimal that emerges is just the
nonlinear version of the LOP. It is also striking that the $(7,1)$
optimal is quantitatively so similar to the full $(23,11)$-optimal
indicating that axial wavenumbers beyond the lowest are not important
for this short pipe calculation. Drastically reducing the azimuthal
resolution to just $(2,1)$ has a noticeable quantitative effect but
still manages to preserve the qualitative features of the NLOP. In
particular the $(2,1)$-optimal (right lower, figure \ref{resgrowth})
captures the essential structure of the $(7,1)$-optimal (right upper,
figure \ref{resgrowth}) which is itself almost identical to the
$(23,11)$-optimal (upper left of figure \ref{NLOPa}) (although no
attempt has been made to match phases of the solutions along the
pipe).

The temporal evolutions of the modal kinetic energies $E_{ml}(t)$
(defined as the kinetic energy associated with the Fourier-Fourier
wavenumbers $m,l$) for the $(7,1)$ calculation are shown in figure
\ref{m7l1} (the equivalent plot for the $(2,1)$ calculation is
qualitatively similar but not shown). The modal energy for
streamwise-independent velocities, $E_{m0}$, is further split into
that associated with the streamwise velocity, $E^w_{m0}$, and that
with the cross-plane velocities $u$ and $ v$, $E_{m0}^{uv}$. Each
modal energy can change because of three effects: input from the
underlying basic state due to the non-normality of the linearised
operator, loss due to viscous dissipation and either loss or gain
through nonlinear mixing with the other modes. Generally, it is
difficult to distinguish between these effects without explicitly
monitoring the various terms in the Navier--Stokes equations. However,
for streamwise-independent modes, the cross-plane energy $E^{uv}_{m0}$
cannot grow by non-normal effects so any energy gain must be the
result of nonlinear input alone. This observation is crucial for
interpreting figure \ref{m7l1}, which shows that after the Orr
mechanism has played out, the NLOP evolution is dominated by the
non-normal energy growth of helical modes ($m,l \neq 0$) in the second
phase ($0.4 \lesssim t\lesssim 2.5$). As these modes grow quickly,
they feed energy via their nonlinear interactions into the
streamwise-independent modes as evidenced in the {\em increase} in
$E^{uv}_{m0}$ over the interval $2 \lesssim t \lesssim 4$. When the
nonnormal energy growth of the helical modes runs out of steam (at
$\approx 2.5$) they decline quickly through the combined effect of
this nonlinear energy drain and viscous dissipation. Thereafter, the
evolution is dominated by each streamwise-independent mode
experiencing slow but sustained non-normal growth as the
secularly-decaying streamwise rolls advect the mean shear to produce
streaks --- the well-known {\em lift-up} process. The uniform decay
rates of the streamwise rolls indicates that there is minimal
nonlinear energy mixing at this point at least in the cross-plane
velocities. This is because they are insensitive to axial advection
(e.g. $w\,\partial u/\partial z=0$) and the cross-plane velocities are
so small.  Figure \ref{resgrowth} shows these two non-normal growth
processes together with the Orr mechanism cooperate nonlinearly to
produce a larger overall growth at $T_{opt}$ than separately. After
the initial rotation and unpacking by the Orr mechanism, the helical
modes grow but quickly run out of steam. They then dump their
energy into the third streamwise-rolls-driving-streaks process, which is
subsequently boosted to reach higher growth factors than otherwise.

%
%
%
%

%
%
\begin{figure}
 \begin{center} \setlength{\unitlength}{1cm} \begin{picture}(12,9.6)
 \put(2  , 4.8){\epsfig{figure=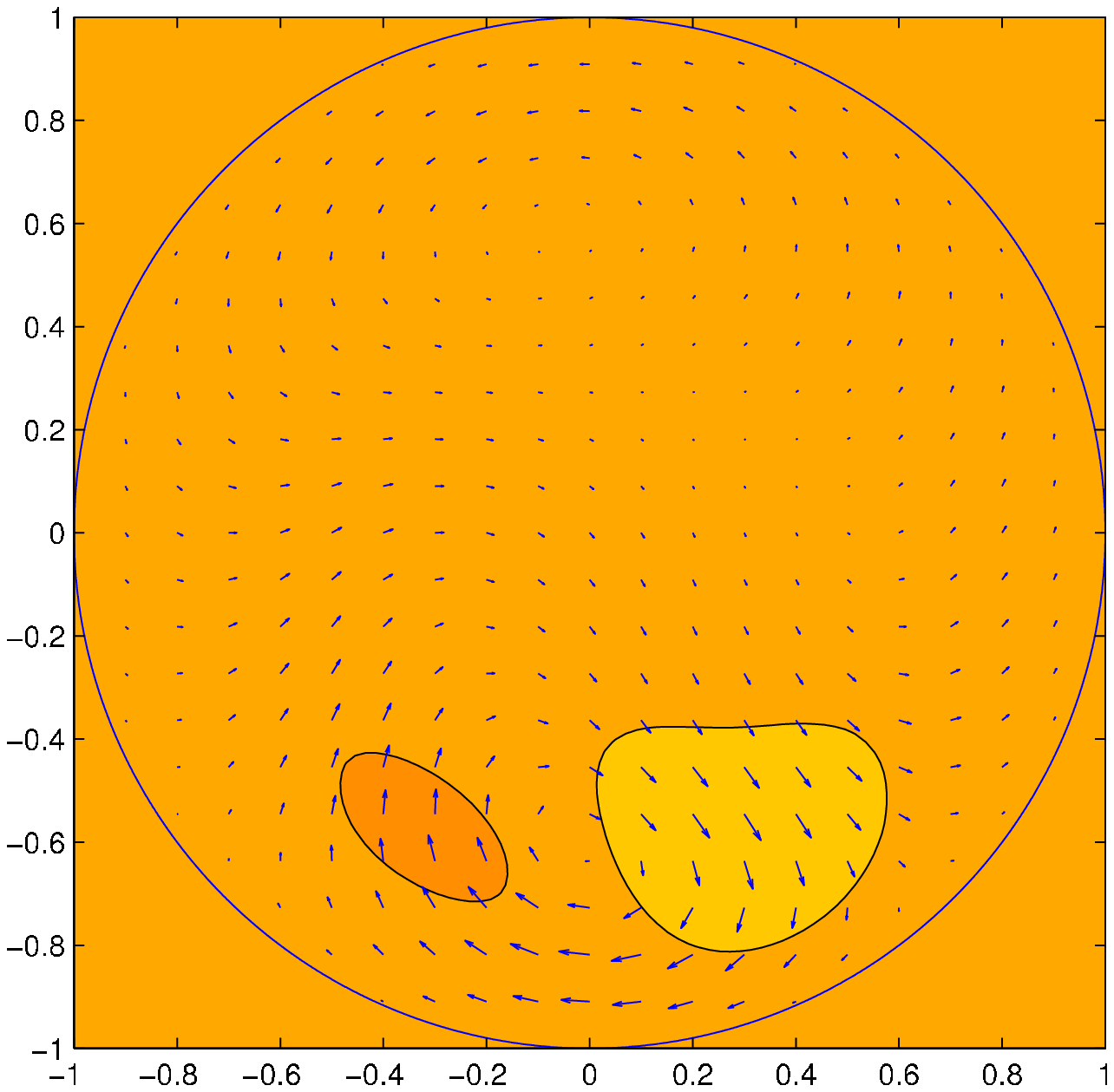,width=4.4cm,height=4.4cm,clip=true}}
 \put(6.6, 4.8){\epsfig{figure=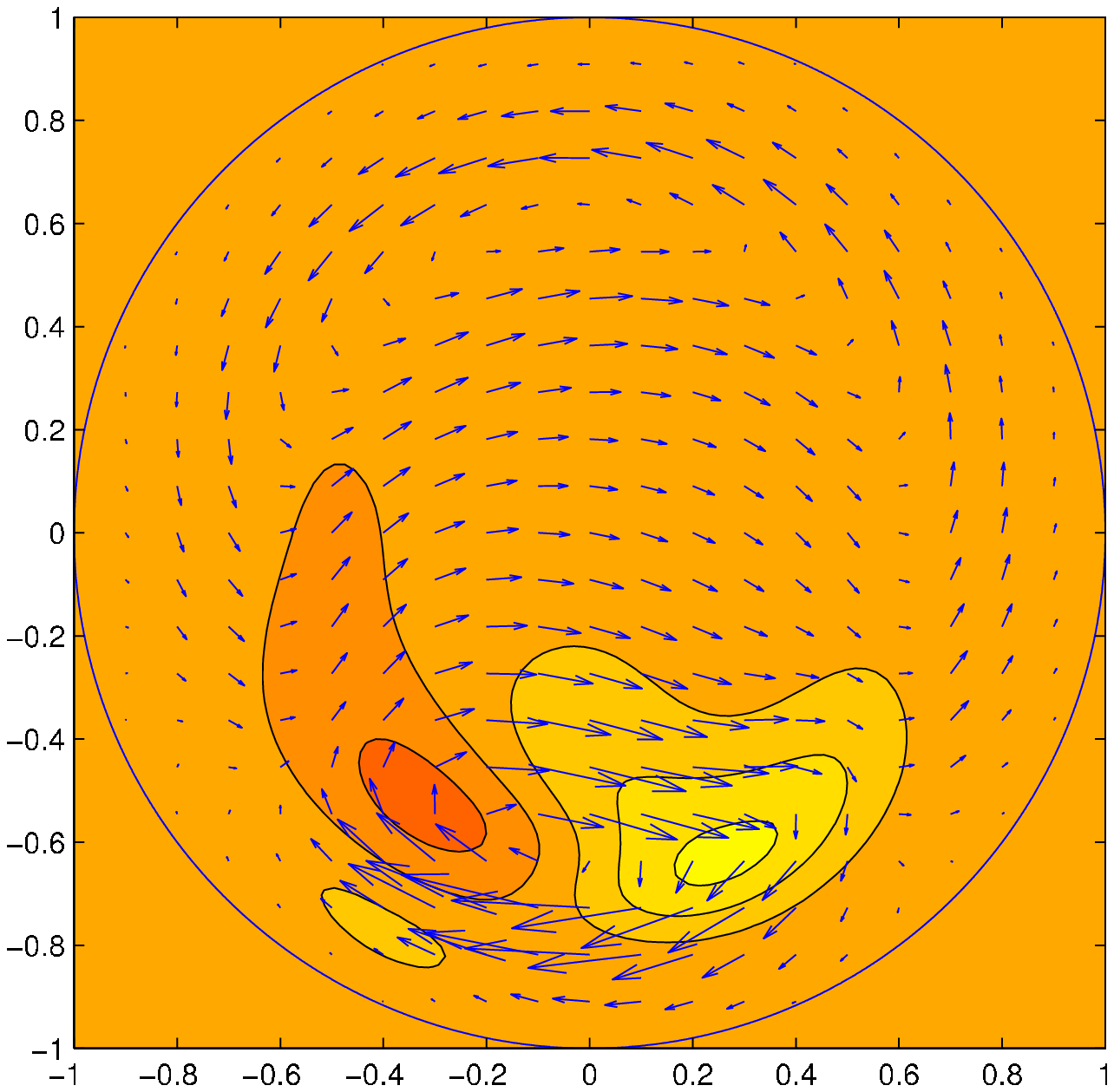,width=4.4cm,height=4.4cm,clip=true}}
 \put(2  ,   0.2){\epsfig{figure=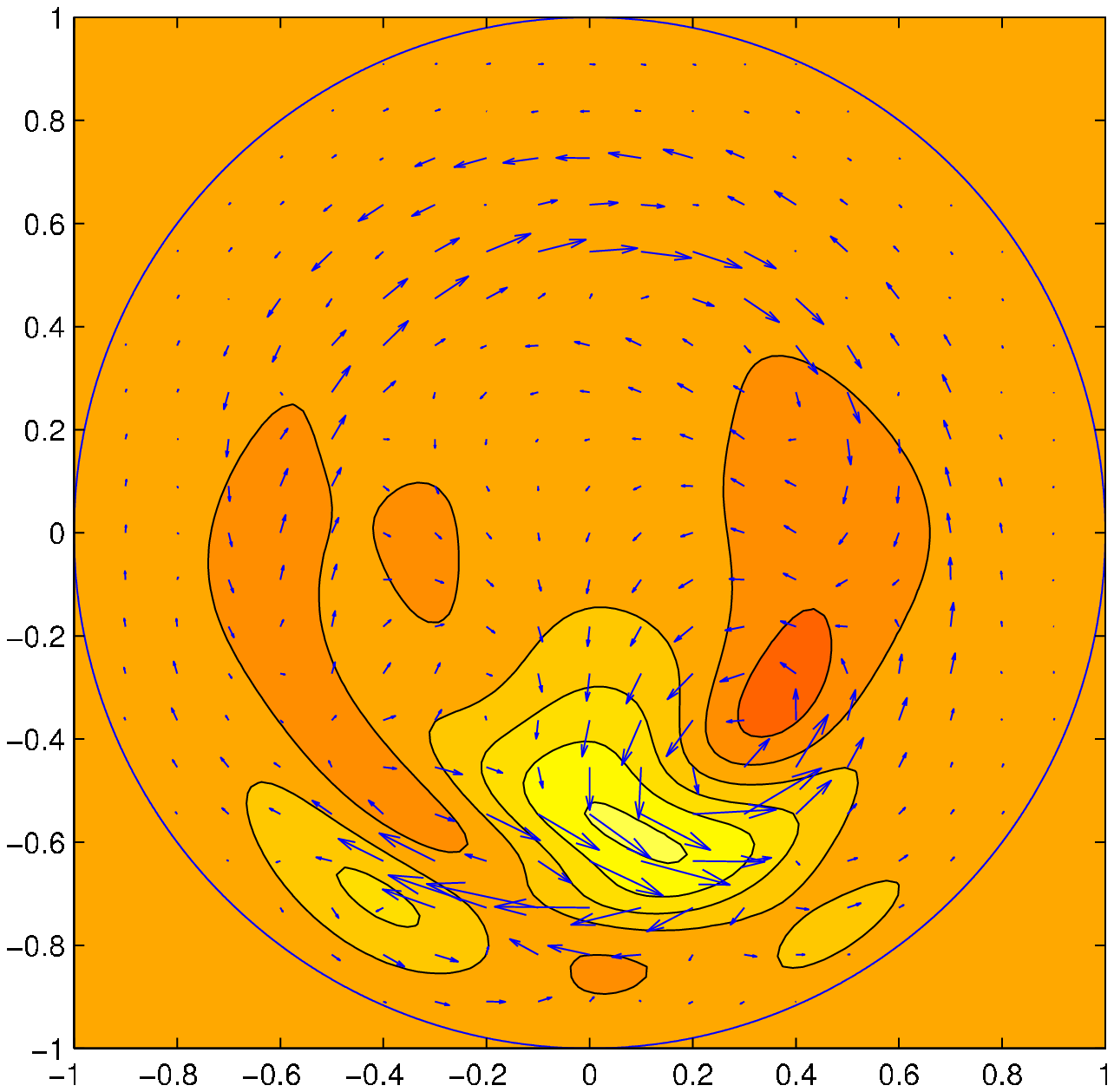,width=4.4cm,height=4.4cm,clip=true}}
 \put(6.6,   0.2){\epsfig{figure=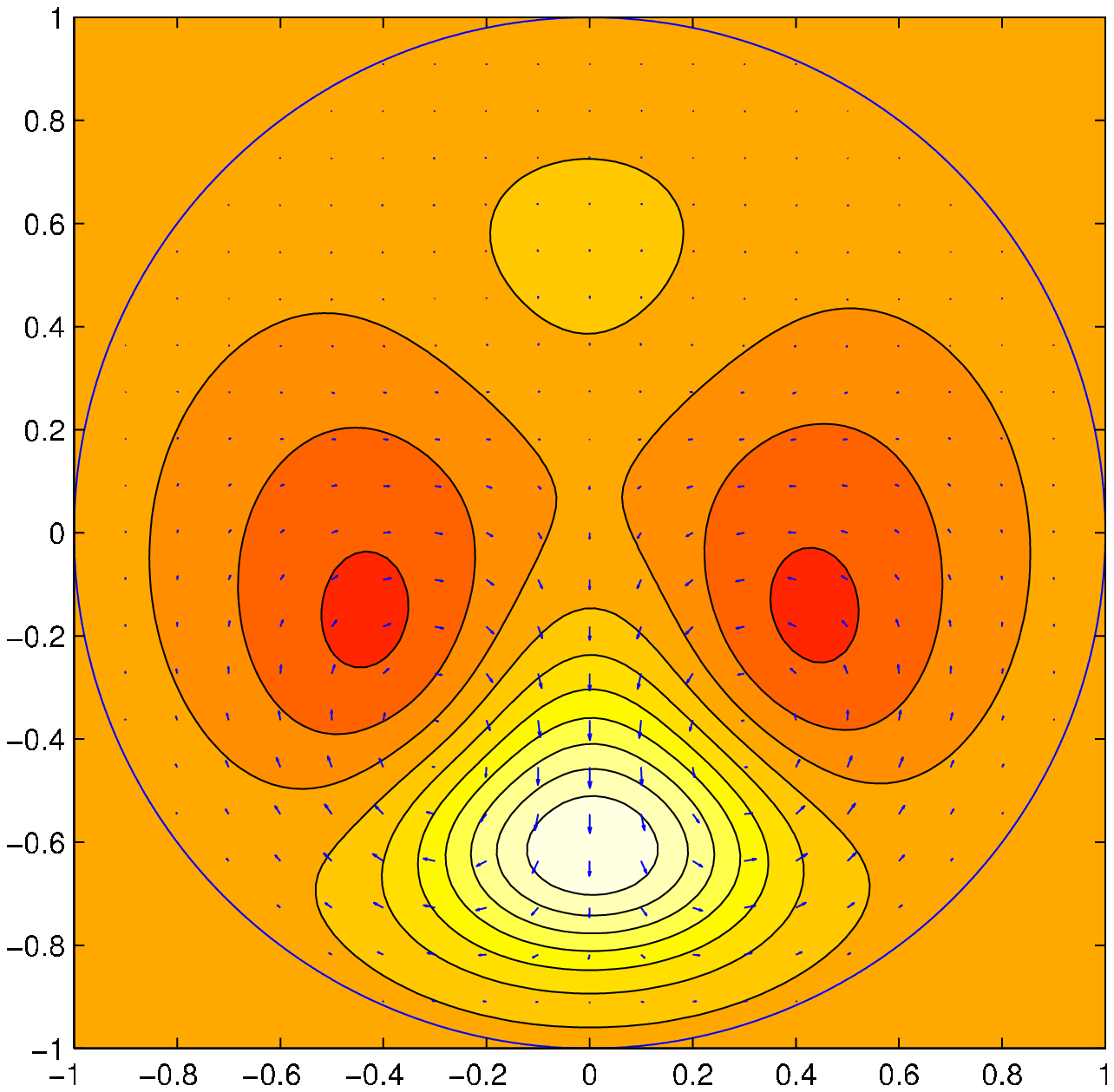,width=4.4cm,height=4.4cm,clip=true}}
 \end{picture} \end{center}
\caption{The NLOP at $Re=1750$ and $E_0=1.8\times 10^{-5}$ calculated
  using resolution (MM,NN,LL)=$(23,64,11)$ at $t=1$ (upper left),
  $t=2.5$ (upper right), $t=4$ (lower left) and $t=10, D/U$ (lower
  right) and representation is as in figure \ref{NLOPa}.  All plots
  have 10 contour levels between the extremes of the streamwise
  velocity perturbation at $t=10$ and cross-stream velocities
  similarly scaled and set by those at $t=2.5$.  Note that the upper
  left plot indicates exactly the same velocity field as the lower
  right plot in figure \ref{NLOPa} but using revised contour and arrow
  levels. }
\label{NLOPb}
\end{figure}

The fact that helical (or more generically `oblique') waves grow best
over short times and streamwise-independent flows grow larger but over
longer times is well known (e.g. figure 8 of Farrell \& Ioannou 1993,
figure 4 of Schmid \& Henningson 1994 and figure 5 of Meseguer \&
Trethen 2003). Furthermore, the scenario of oblique waves growing
transiently, feeding their energy into streamwise rolls that then
drive streamwise streaks (which then become unstable) has also been
proposed before as an efficient bypass mechanism in Reddy {\em et
al.}\ (1998) (called the `oblique wave scenario'). This general picture,
or at least the first stages of it, appear to be confirmed here in the
nonlinear growth problem. However, the initial localisation of the
perturbation and how it `unwraps' to give a final, large,
predominantly streamwise-independent flow is a new feature born out of
a need to cheat the starting (global) energy constraint. Figures
\ref{NLOPa} and \ref{NLOPb} show how the structure of the NLOP across
one (fixed) slice of the pipe evolves in time. The initial slice shown
(upper left and again upper middle but rescaled) has a peak
cross-plane speed of $\approx 0.02U$ concentrated in a tight vortex
pair near the pipe wall and peak axial speed of $\approx 0.012U$. The
delocalisation or `unwrapping' is effectively completed by $t \approx
1$ when the peak cross-plane speed is essentially unchanged whereas
the peak axial speed has grown to $0.06U$. Figure \ref{NLOPb} (the
upper left slice is a rescaled version of the lower right of figure
\ref{NLOPa}) shows that {\em both} the cross-plane and axial speeds
grow considerably in the interval $1 \lesssim t \lesssim 2.5$ (peak
cross-plane speed increases from $0.02U$ to $0.08U$ and peak axial
speed from $0.06U$ to $ 0.13U$). In fact, by $t=2.5$ the initial
energy has experienced most of its growth (a factor of $\approx 50$)
and only a further magnification by $\approx 7$ follows in the next
$\approx 20\, D/U$. In this latter period the cross-plane velocities
manoeuvre the streak structure into place and then die away so that
even by $t=10$, the predominantly streamwise-independent and axial
flow has been established (peak cross-plane speed is $0.012U$ and peak
axial speed $0.34U$ now). It is worth stressing that even at this
point, the flow does not match the LOP (see plot $c^{'}$ in figure 2
of PK10), which depends solely on the Fourier-Fourier basis function
$\exp(\mathrm{i}\phi)$ and is strictly 2D, being streamwise-invariant.  

\section{Tracking the NLOP in PK10\label{sec:PK10}}

As discussed in section \ref{sec:intro}, PK10 demonstrated how
nonlinearity can qualitatively change the form of the optimal
disturbance of a given initial energy which achieves the most energy
growth over a fixed period. The new NLOP could not, however, be
followed up to the initial energy level at which turbulence was
triggered. We now revisit this situation armed with a more efficient
and parallel code which allows higher resolution and more carefully
refined steps in $\boldsymbol{\bu}_0$  to be used.

\begin{figure}
\centering
\resizebox{0.9\textwidth}{!}{\includegraphics[angle=270]{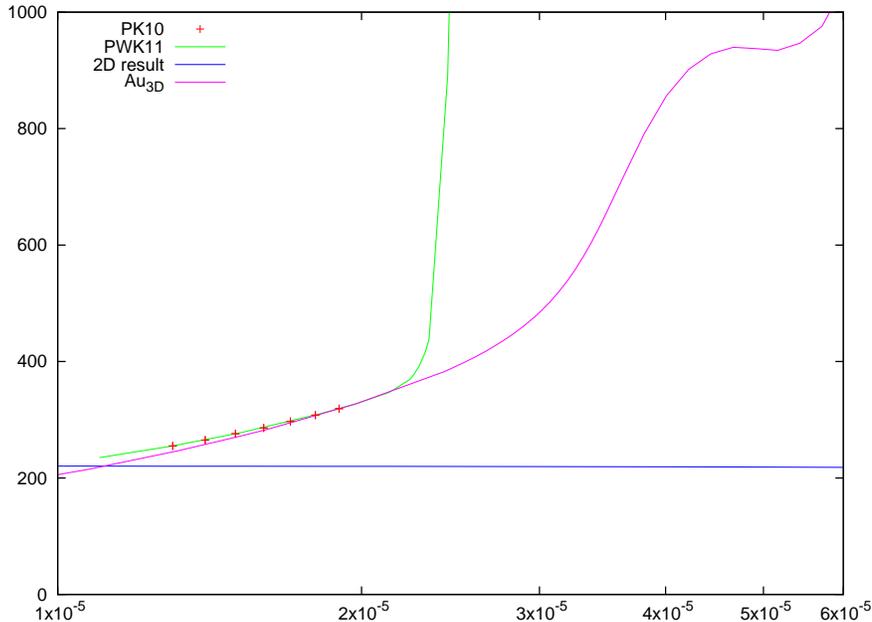}}
\caption{Reproduction of figure 5 from Pringle \& Kerswell (2010)
  showing growth as a function of initial energy. The red crosses
  correspond to the nonlinear optimal perturbations previously
  calculated in PK10, while the (uppermost) green line represents the
  optimals calculated with the new code presented in this paper. The
  (middle) magenta line shows the amount of growth produced by simply
  rescaling the nonlinear optimal for $E_0=2\times10^{-5}$ and
  therefore lower bounds the true (green) optimal growth curve. The
  (lowest) flat blue line is the growth provided by the nonlinearly
  modified 2D optimal (LOP).  }
\label{fig:g1750}
\end{figure}

In PK10, a short $\half \pi D$ periodic pipe was adopted to minimise
the axial resolution needed and the relatively short time period was
taken equal to $T_{lin}$, the time for maximum energy growth in the
linearised Navier-Stokes equations, to highlight the effect of
nonlinearity. Working at $Re=1750$, PK10 report failing to converge
for $E_0>2\times 10^{-5}$.  Their best estimate for $E_c$ was
$\mathscr{E}_c = 6 \times 10^{-5}$, the energy required to trigger
turbulence when using a perturbation of the form $A\bu_{3d}
(\bx;E_0=2\times 10^{-5}, Re=1750)$. With the new code using a
resolution $(MM,LL)=(23,11)$ and $NN=64$ finite difference radial
points as opposed to PK10's $(MM,LL)=(14,5)$ and a 25 Chebyshev
polynomial expansion radially, we were able to confirm PK10's results
for $E_0 \leq 2 \times 10^{-5}$ as well as continuing to converge up
to $E_0=2.52 \times 10^{-5}$. Above this point, the amount of growth
grows sharply compared with the amount of growth produced by rescaling
arguments (figure \ref{fig:g1750}).

%
%

\begin{figure}
\centering
\resizebox{0.9\textwidth}{!}{\includegraphics[angle=0]{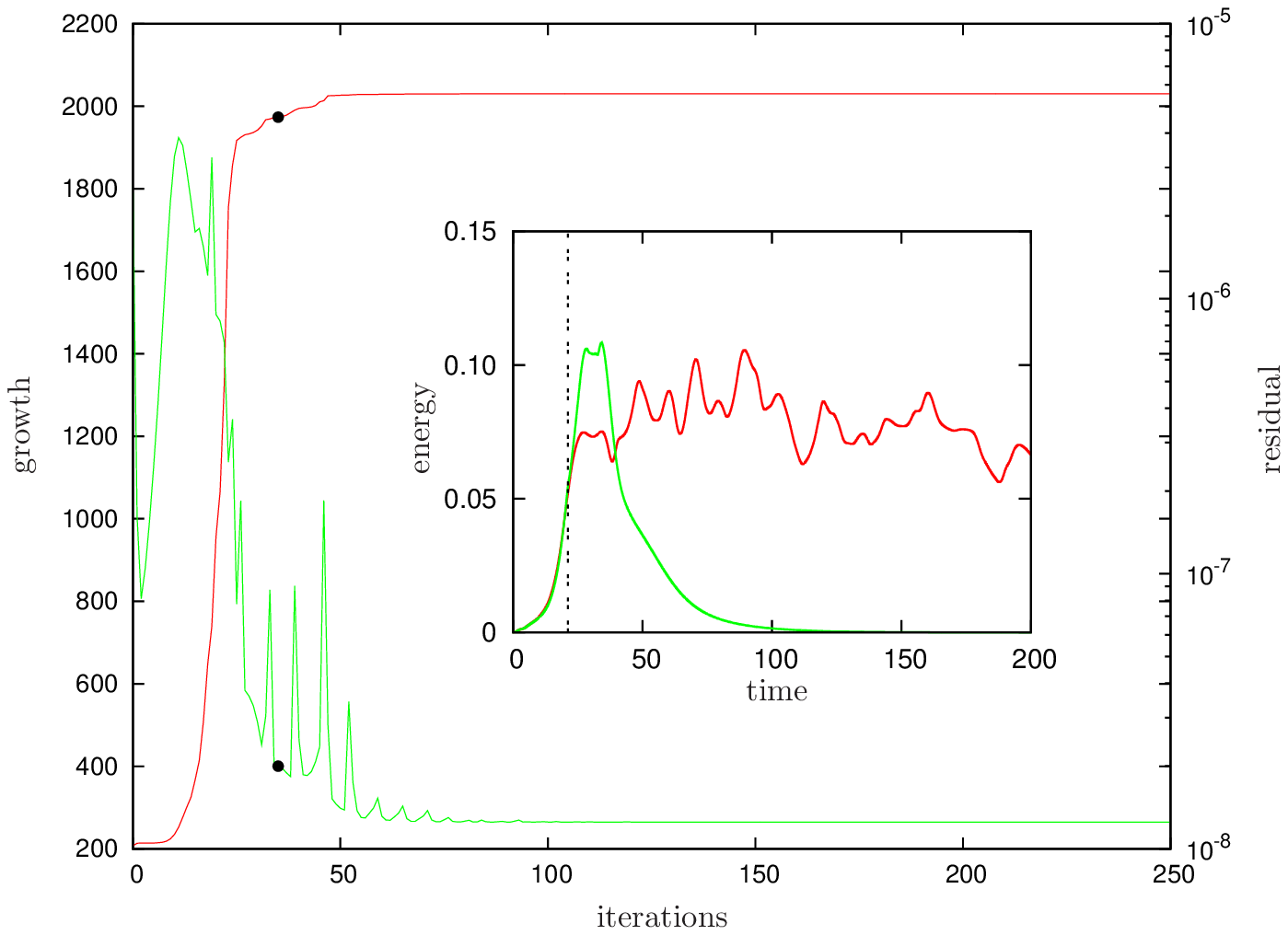}}
\caption{Convergence of procedure at $E_0=2.52\times 10^{-5}$.  The
  algorithm smoothly converges to a growth of 2030 (upper right line),
  while the residual decays to $O(10^{-8})$ (lower right line). The
  algorithm was continued for over 800 iterations in total to ensure
  that there was no further change. \textbf{Inset}: The time
  evolutions of two disturbances found by this sequence of
  iteration. The green (lower right) line is the converged optimal
  which smoothly relaminarises, while the (upper right) red line is
  the disturbance corresponding to the black dot leading to a
  turbulent episode. The vertical dashed line shows the target time
  $T_{opt}=T_{lin} \approx 21.3\, D/U$ from the optimisation
  procedure.  }
\label{fig:conv1750}
\end{figure}

Examining how the residual $\langle
(\delta\mathscr{L}/\delta\vec{u}_0)^2\rangle$ decreases as the
algorithm proceeds indicates that the code has converged at
$E_0=2.52\times 10^{-5}$: see figure \ref{fig:conv1750} (outer). The
time evolution of the optimal solution is relatively smooth,
relaminarising after the initial transient growth. Surprisingly,
however, one of the velocity fields it iterates through (marked with a
black dot in figure \ref{fig:conv1750}, outer) \emph{does} lead to a
turbulent episode. A comparison of the two evolutions confirms the the
optimisation procedure has worked properly: the optimal produces more
growth than the other initial condition despite the fact it doesn't
lead to turbulence (figure \ref{fig:conv1750}, inner). This
observation seems to go against our assertion that the optimisation
algorithm will latch onto a turbulence-triggering state and then fail
to converge. There are two important lessons to be learnt from this
apparent pathology. The first, most obvious one is that the
turbulence-triggering initial condition has not had enough time to
reach the turbulent state by the end of the (short) period $T_{lin}$,
so {\bf (1) $T_{opt}$ needs to be large enough}. Secondly, figure
\ref{fig:conv1750} shows that the energy level reached by the optimal
a little after $T_{lin}$ is actually higher than that typically
associated with the turbulent state at this (low) $Re$ in this (tight)
pipe geometry. This situation is fatal for the approach being
advocated here, which relies on the turbulent state producing the
highest values of the energy growth (or whatever functional is being
considered) in comparison to non-turbulent states. Fortunately, such a
situation only seems to occur in tightly-constrained (small geometry)
flows close to (in $Re$) the first appearance of the turbulent
state. 
Therefore, the second lesson is that {\bf (2) the optimisation
  strategy can only be used sufficiently far from the first appearance
  of turbulence and/or for flows in large domains}.
 
In hindsight then, the geometry and $Re$ value chosen in PK10 is not
suited for determining $E_c$ there using this optimisation
approach. We therefore switch to a longer $5D$ periodic pipe
(theoretically popular since the work of Eggels {\em et al.}\ 1994) and a
higher $Re=2400$ where the edge shows typical behaviour (Duguet,
Willis \& Kerswell 2008) and the turbulent state is clearly
energetically separated from the edge state (e.g.\ Schneider \&
Eckhardt 2009, figure 7).

%
%

%
%
\begin{figure}
\centering
\resizebox{0.75\textwidth}{!}{\includegraphics[angle=0]{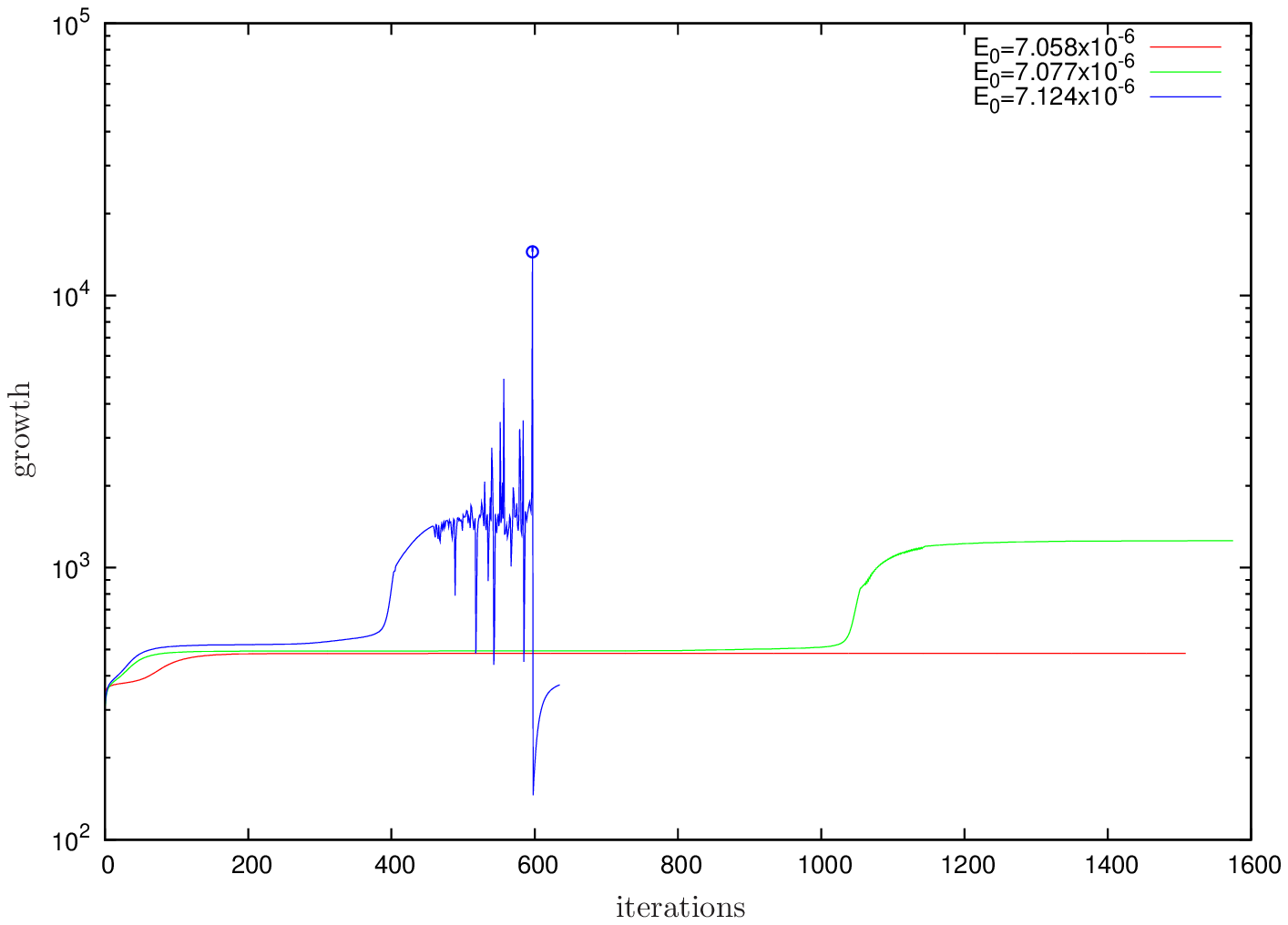}}
\caption{The amount of growth produced by successive choices for
  $\bu_0$ in the interative scheme. For $E_0=7.058\times 10^{-6}$ the
  growth plateaus out at 483 (red lower line). For $E_0=7.124\times
  10^{-6}$, the growth briefly plateaus at $\sim 520$ before rapidly
  rising to excess of $1,000$ (upper blue line). The circle plotted
  shows a growth of $14,480$ and has reached the turbulent attractor
  before the algorithm steps momentarily back to a region below the
  edge. The turbulent seed and the optimal for $E_0=7.058\times
  10^{-6}$ are both plotted in figure \ref{fig:minSeed}. Before this
  has been reached, however, the growth being produced is already
  nonsmooth between steps due to the lack of smoothness in the
  hypersurface $\mathscr{L}$, as predicted by conjecture 1. We also
  include the choice of $E_0=7.077\times 10^{-6}$ for illustrative
  purposes (middle green line). This iterative run has not been fully
  converged and it is not clear whether it will converge to an optimal
  or depart to the turbulent state.}
\label{fig:growth2400}
\end{figure}

\section{Long Time and a $5\,D$ Pipe\label{sec:longTime}}

In order to assess the twin conjectures discussed in section
\ref{sec:intro}, a practical decision needs to be made as to what
constitutes a `asymptotically long' optimisation time. Figure
\ref{fig:conv1750} shows that an initial condition is capable of
growing through several orders of magnitude into a turbulent epsiode
within $\sim 50\,D/U$. We therefore chose $T_{opt}=75\,D/U$, which
should be large enough to capture this behaviour especially in a
larger $5D$ domain although $Re=2400$ is higher (so
$T_{opt}>2.5T_{lin}$ at this $Re$). It is worth remarking, though,
that this finite choice will limit the accuracy to which we can
determine the energy threshold. The algorithm senses initial
conditions which have reached the turbulent state by the end of the
observational window. This sets a lower limit on how close they can be
to the edge, as the time for a turbulence-triggering initial
condition to reach turbulence becomes arbitrarily large as it is taken
closer to the edge. This said, our choice of $T_{opt}$ gives
acceptable accuracy yet the way to improve this is clear through
integrating for longer.

The results of the energy growth optimisation in a $5\,D$ pipe at
$Re=2400$ as a function of $E_0$\footnote{Note that the
    nondimensionalisation of energy is dependent on the size of the
    flow domain being considered, and so equivalent absolute energies
    will appear smaller after nondimensionalising in this longer
    pipe.} exactly mimicks the situation uncovered in PK10.  For
$E_0$ small enough, the linear (streamwise-independent) optimal (LOP)
is selected albeit with slight nonlinear modification, which
suppresses the growth of the 2D optimal as $E_0$ increases. 
Then there is a finite value (PK10
refer to this as $E_{3d}$) when a new 3D optimal (NLOP) is
preferentially selected, which shows localisation in the azimuthal and
radial directions. There is also some localisation in the streamwise
direction, however the domain is by no means long enough for us to
observe truly localised optimals as opposed to periodic disturbances.

\begin{figure}
\centering
\resizebox{0.75\textwidth}{!}{\includegraphics[angle=0]{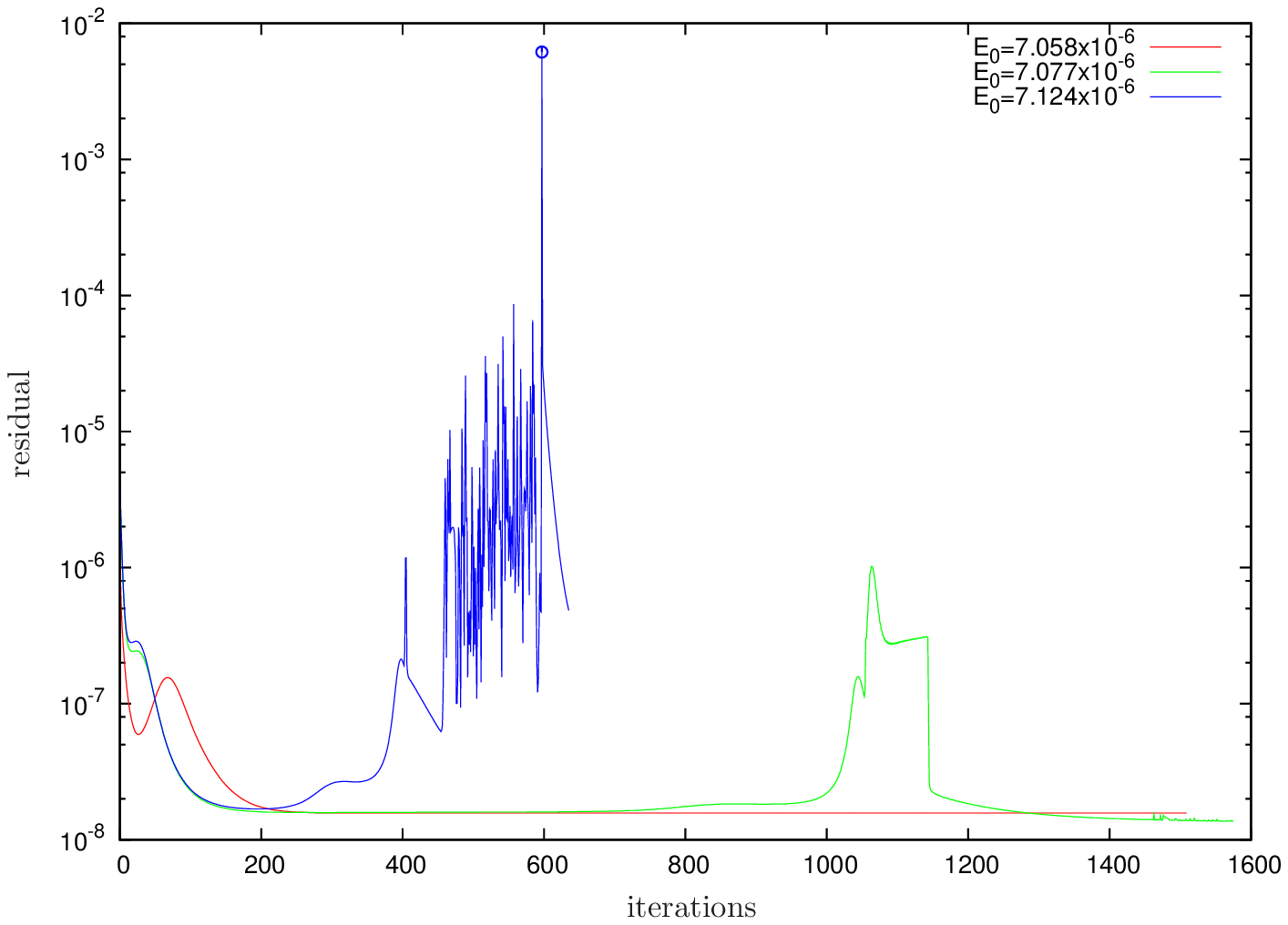}}
\caption{The residuals corresponding to the three iterative runs
  described in figure \ref{fig:growth2400} (flat red line $7.058 \times
  10^{-6}$, green line with spike at $1100$ $7.077 \times 10^{-6}$ and
  upper blue $7.124 \times 10^{-6}$). The sudden adjustment centred on
  1100 iterations is a warning that deciding upon convergence can be a
  subtle affair.}
\label{fig:res2400}
\end{figure}

\begin{figure}
\centering
\resizebox{0.75\textwidth}{!}{\includegraphics[angle=0]{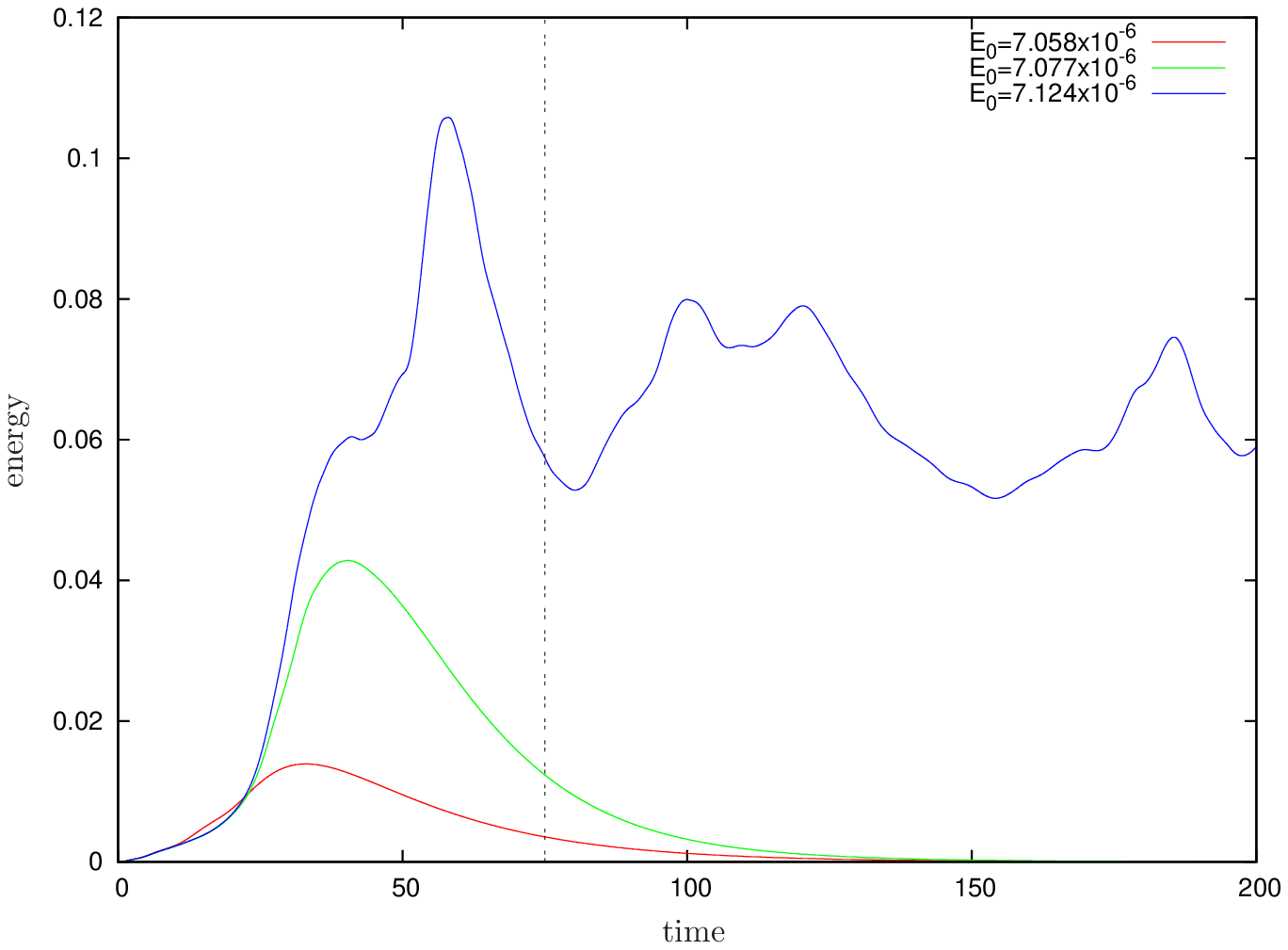}}
\caption{Evolution of the final states produced by the iterative
  scheme for $E_0=7.058\times 10^{-6}$ (red lowest line) and
  $7.124\times 10^{-6}$ (blue uppermost line).  One clearly leads to a
  turbulent episode while the other simply relaminarises after the
  intial transient growth. Also shown is the evolution of the initial
  condition arrived at after 1600 iterations for $E_0=7.077\times
  10^{-6}$ (middle green line) which again relaminarises. The vertical
  line marks the optimisation time.}
\label{fig:evol2400}
\end{figure}

As $E_0$ is increased further, there comes a point at which the
algorithm struggles to converge properly. Successive bisection
indicates that this value, $E_{fail}$, is bracketed by the initial
energy values of $E_0=7.058\times 10^{-6}$ which converges smoothly to
the NLOP and $7.124\times 10^{-6}$ which clearly fails due to the
occurrence of a turbulence-triggering initial condition: see figures
\ref{fig:growth2400}, \ref{fig:res2400} and \ref{fig:evol2400}.

%
%

%
%

An attempt to improve this bracketing by taking $E_0=7.077\times
10^{-6}$ appears to show convergence yet there still remains some
doubt even after running the algorithm for nearly 1600 iterations and
50,000 CPU hours ($\approx 6$ years). Figure \ref{fig:growth2400} indicates
convergence yet at a much higher level compared to that reached by the
`nearby' initial energy of $E_0=7.058\times 10^{-6}$. Moreover the
fact that there is a jump up to this higher level after $\approx 1000$
iterations is mildly disconcerting. This adjustment is reflected in
the evolution of the residual (see figure \ref{fig:res2400}), which
after $200$ interations, appears to show convergence for the next $500$
iterations before being followed by a rapid
transition that ends after $1150$ iterations. It is not clear
whether the algorithm now has finally converged or whether it will
subsquently encounter a turbulence-inducing initial condition. This
example demonstrates that it is clearly very important to take care
when deciding whether the procedure has converged or not
(e.g. stopping the algorithm after 600 iterations would indicate clear
convergence). It is probable that the algorithm is struggling to
discern between turbulence-inducing initial conditions and the NLOP
because the time to reach turbulence is comparable to, or exceeds,
$T_{opt}$. Consequently, the estimate that $7.058\times 10^{-6} <
E_{fail} < 7.124\times 10^{-6}$ is the best we can hope for working
with $T_{opt}=75\, D/U$ and the fate of $E_0=7.077\times 10^{-6}$
could be decided by taking a longer $T_{opt}$ (not pursued here).

%
%
\begin{figure}
\centering
\resizebox{1\textwidth}{!}{\includegraphics[angle=0]{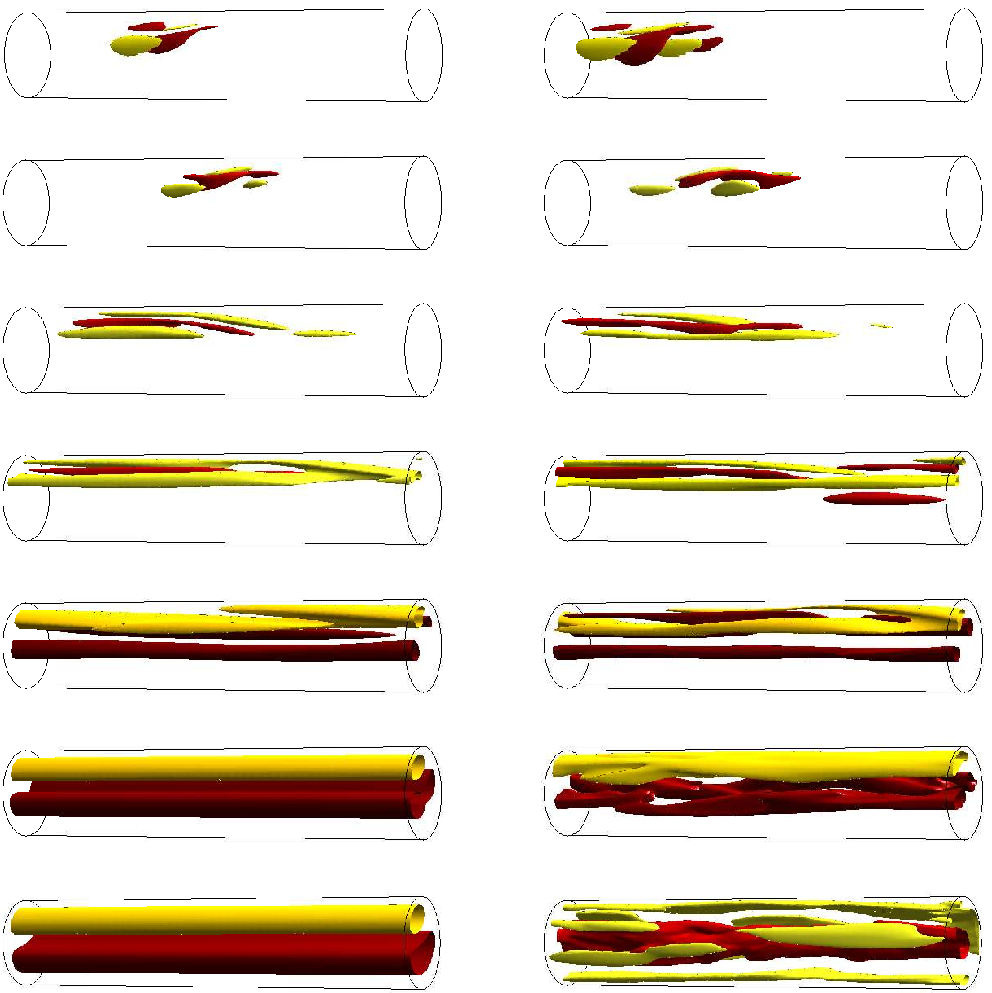}}
\caption{Snap shots showing isocontours of streamwise perturbation
  velocity during the evolution of the final states produced by the
  iterative scheme for $E_0=7.058\times 10^{-6}$ (left) and
  $7.124\times 10^{-6}$ (right). The isocontours in each plot
  correspond to $50\%$ of the maximum (light yellow) and $50\%$ of the
  minimum (dark red) of the streamwise perturbation velocity in the
  pipe \emph{at that time}. The snapshots correspond to times $t=0$,
  $0.5$, $5$, $10$, $20$, $40$ and $75\,D/U$. In both cases the energy
  is initially localised in the streamwise direction and the
  disturbance quickly spreads. By $t=10$ both disturbances have
  created streamwise streaks but only for the lower energy do they
  become streamwise independent.  The larger amplitude of the higher
  energy streaks are subject to a turbulence-triggering
  instability.}
\label{fig:5dIso}
\end{figure}

\begin{figure}
\centering
\resizebox{0.9\textwidth}{!}{\includegraphics[angle=0]{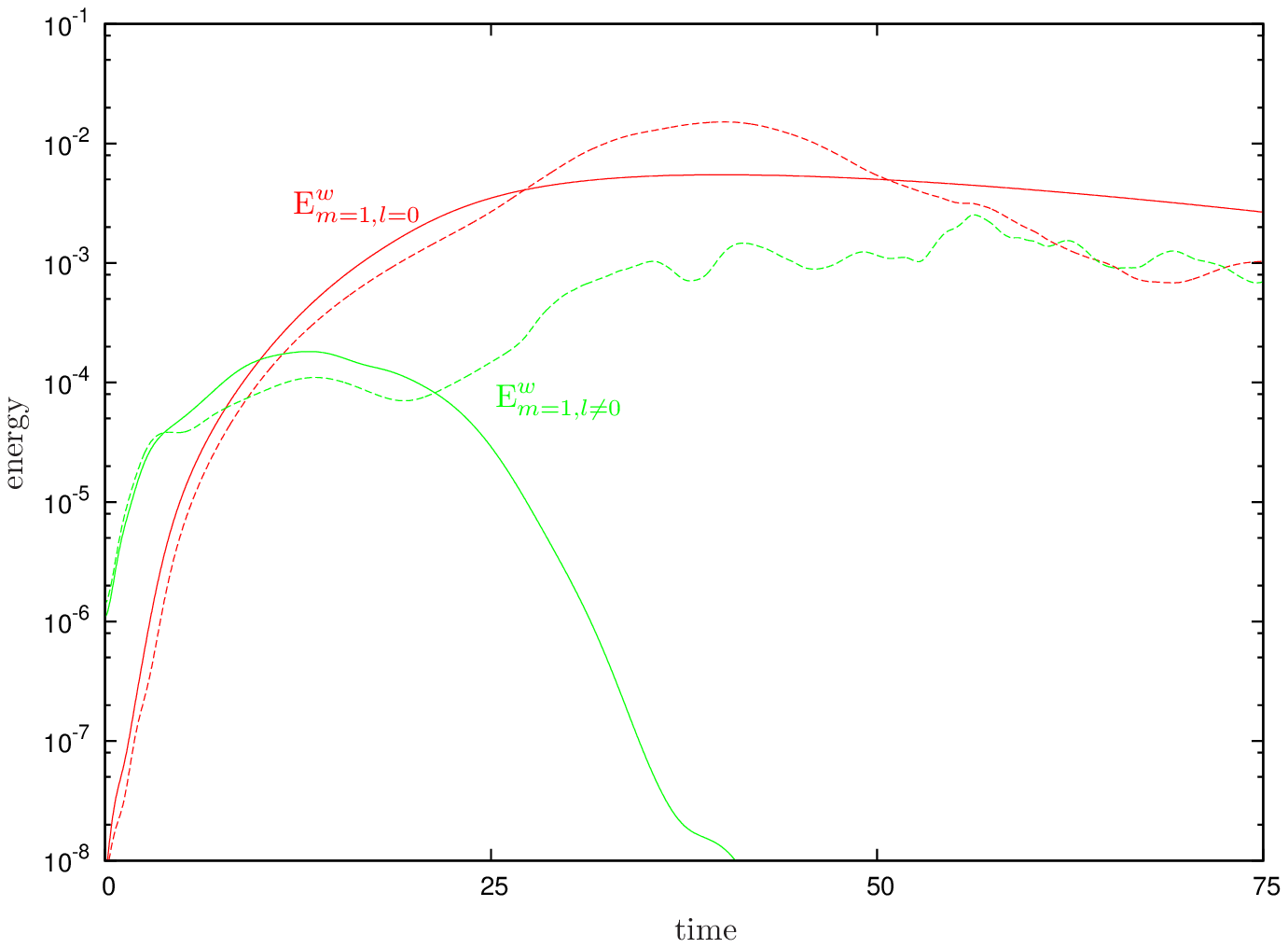}}
\caption{The energy associated with the $m=1$ axial component of the
  disturbances calculated for $E_0=7.058\times 10^{-6}$ (solid) and
  $E_0=7.124\times 10^{-6}$ (dashed). Each energy is split into
  streamwise-independent (dark red uppermost lines at $t=25D/U$) and
  streamwise-dependent (light green lowermost lines at $t=25D/U$)
  parts. The former measures the streaks created by the disturbance
  while the latter shows the instability of these streaks.}
\label{fig:2400modes}
\end{figure}

The physical evolution of the two disturbances is shown in figure
\ref{fig:5dIso}. Initially the disturbances look streamwise-localised
because of the contouring but they do in fact occupy the full length
of the domain. Both subsequently develop into coherent
domain-length streaks. In the $E_0=7.058\times 10^{-6}$ case, these
streaks continue to evolve yet remain stable, becoming almost totally
streamwise-invariant before ultimately decaying. In the
$E_0=7.124\times 10^{-6}$ case, the streaks have higher amplitude and
a streak instability clearly occurs leading to turbulence. The nature
of this instability is shown in figure \ref{fig:2400modes} which plots
the streamwise dependent and independent components of the energy of
the axial velocity field with (lead) azimuthal wave number $m=1$. For
the relaminarising disturbance, the 3D part of the energy decays
monotonically from around $15\,D/U$ onwards.  For the more energetic
disturbance the decay is abated after $20\,D/U$ at which point, an
instability of the streaks occurs eventually leading to turbulence. It
is worth remarking that the final plot for $E_0=7.058\times 10^{-6}$
in figure \ref{fig:5dIso} resembles more the two streaks produced by
the linear optimal, rather than the three-streak field generated in
PK10 (cf figure \ref{NLOPb}, bottom right in this paper). Whether this
is due to the increased target time or the lengthened flow domain is
unclear.

%
%

\begin{figure}
\centering
\resizebox{0.8\textwidth}{!}{\includegraphics[angle=270]{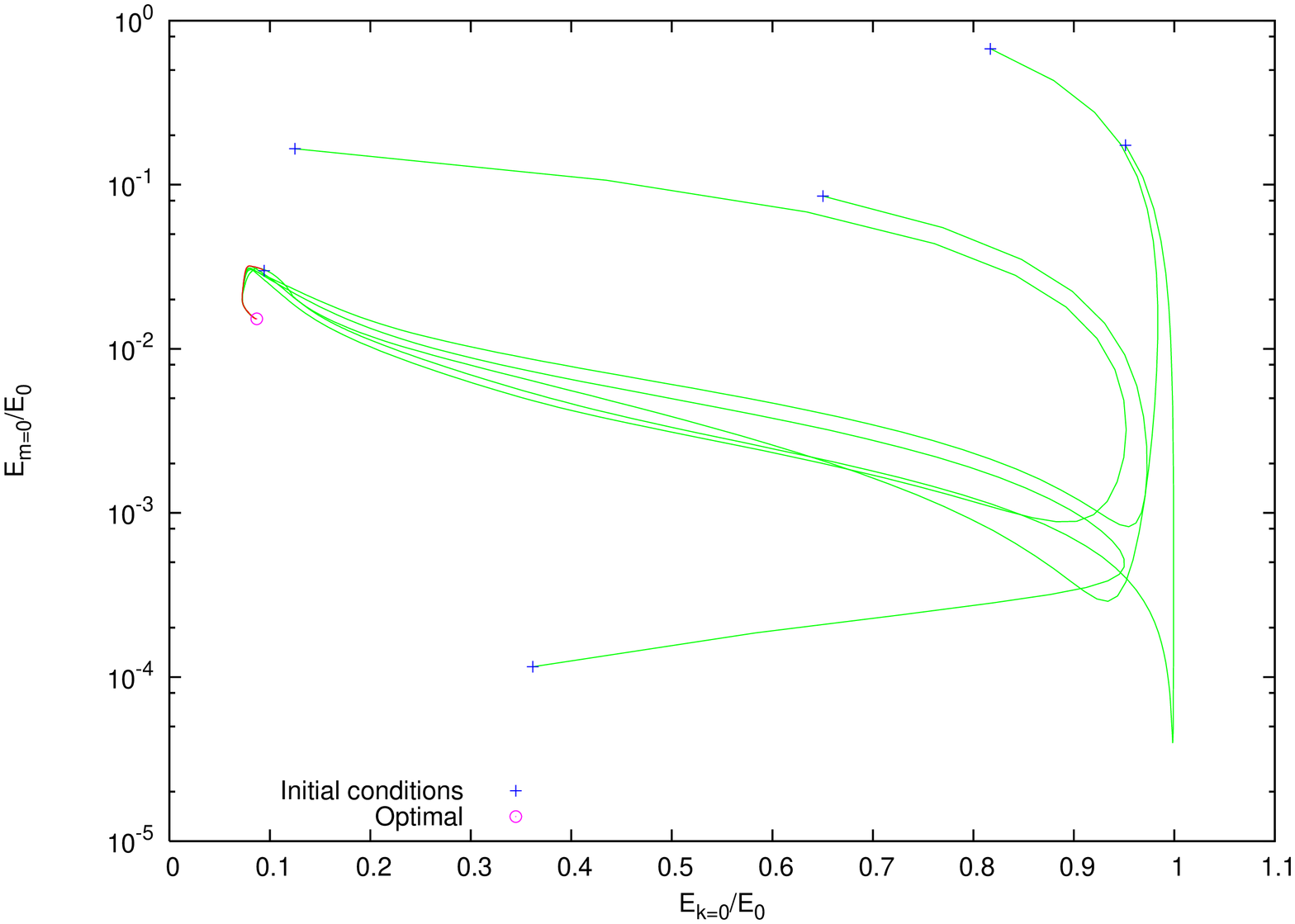}}
\caption{Convergence of six different initial conditions (crosses)
  towards the same optimal (circle) for $E_0= 7.058\times 10^{-6}$:
  axes are normalised perturbation energies associated with the
  streamwise-independent part (abscissa) and the axisymetric part
  (ordinate).  The initial conditions chosen correspond to the various
  combinations of turbulent flow fields, travelling wave solutions and
  the nonlinear optimal from section \ref{sec:PK10}. The red line
  corresponds to this final choice and is the iterative scheme shown
  in figures \ref{fig:growth2400} and \ref{fig:res2400}.}
\label{fig:phaseConv1}
\end{figure}

\begin{figure}
\centering
\resizebox{0.8\textwidth}{!}{\includegraphics[angle=270]{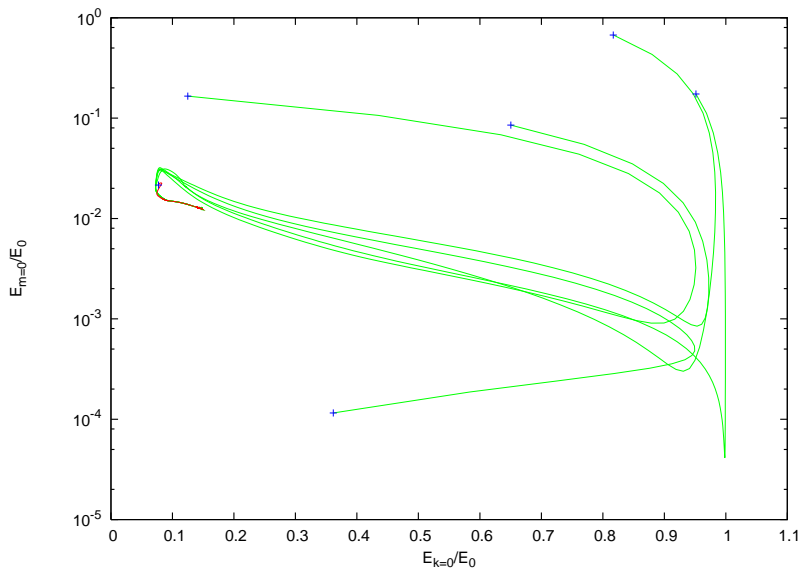}}
\caption{ The result of using the same six initial conditions
  (crosses) for $E_0= 7.124\times 10^{-6}$. The red line corresponds
  to the iterative progression shown in figures \ref{fig:growth2400}
  and \ref{fig:res2400}. Clearly, the procedure is independent of the
  starting guess. Note also how similar the progression is here to
  that for $E_0=7.058\times 10^{-6}$.  }
\label{fig:phaseConv2}
\end{figure}

Conjecture 1 claims $E_c =E_{fail}$.  As there is no evidence of
turbulence-triggering initial conditions at $E_0=7.058\times 10^{-6}$
but there is at $7.124\times 10^{-6}$, we have 
$7.058\times 10^{-6} < E_{c} < 7.124\times 10^{-6}$.  To this
level of accuracy we have found that $E_c=E_{fail}$. 
That the optimisation scheme
will fail if turbulent seeds exist within the $E_0$-hypersurface seems
clear \emph{provided the iterative scheme can find them}. Establishing
this is very difficult if not impossible, but a weaker practical
alternative is to demonstrate that the procedure is not dependent on
the initial starting guess $\bu_0$. To do this we have compared six
very different choices for the initial seed for both $E_0<E_{fail}$
and $E_0>E_{fail}$ and plotted their evolution on a 2D projection of
energy in the axisymmetric part of the perturbation against energy in
the streamwise-independent part (figures \ref{fig:phaseConv1} and
\ref{fig:phaseConv2}). The scatter of the initial crosses illustrates
the variety of initial conditions used which range from turbulent
velocity fields to known travelling wave solutions. In both cases,
irrespective of where the scheme begins, the eventual (iterative)
evolution brings it to the same trajectory in this `phase space'. This
provides some evidence that the algorithm does sample the
$E_0$-hypersurface well and that Conjecture 1 indeed holds true.

%
%

%
%

%
%
\begin{figure}
\centering
\resizebox{1\textwidth}{!}{\includegraphics{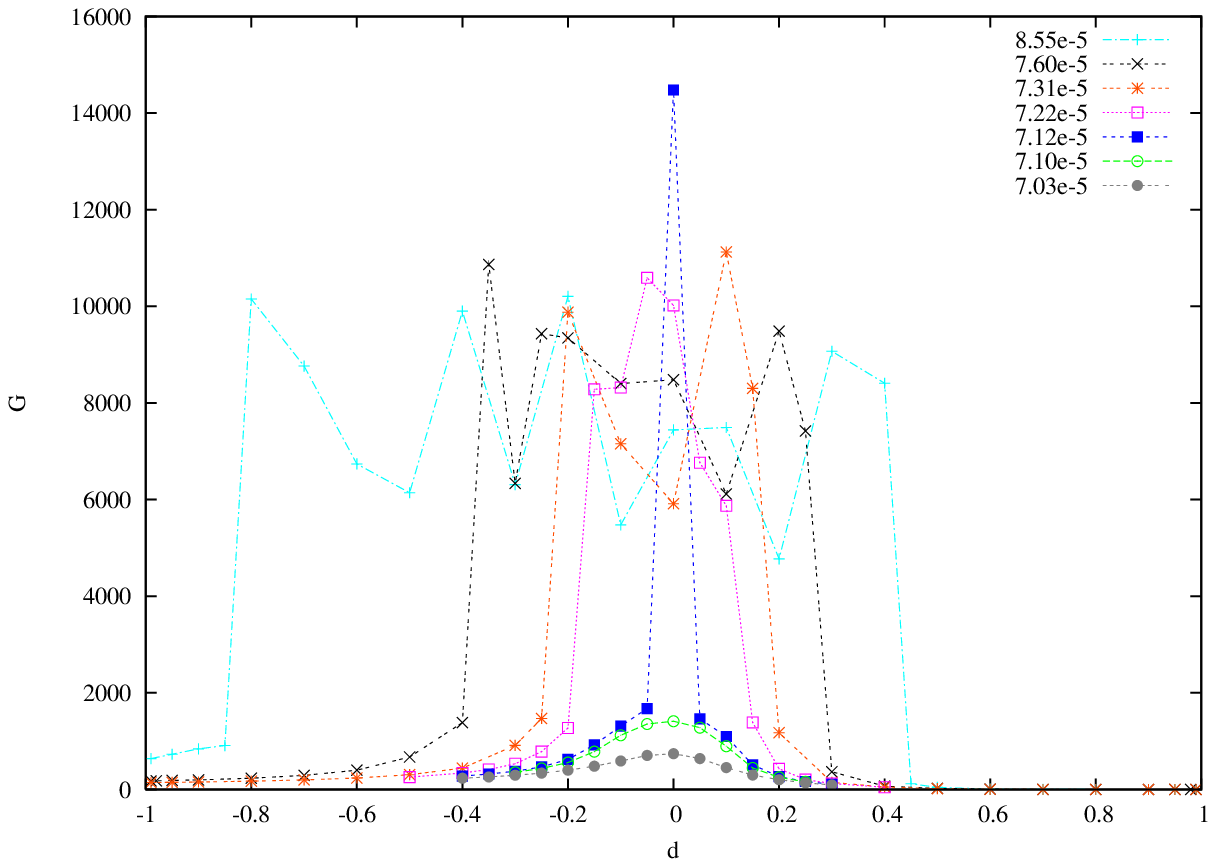}}
\caption{Growth factors in the neighbourhood of the seed state $\bu_t$
  that triggers turbulence at $E_0=7.124 \times 10^{-6}$.  The
  neighbourhood is defined by $d$ where
  $\bu_{ic}(\bx;d):=A[\,(1+d)\bu_{l=0}+(1-d)\bu_{l\ne 0}\,]$, $\bu_{l=0}$ is
  the streamwise-independent part of $\bu_s$ and
  $\bu_{l \ne 0}:=\bu_s-\bu_{l=0}$. The amplitude $A$ is used to rescale the
  initial state to ensure the correct starting energy $E_0$. The
  figure shows that $\bu_s$ only just triggers transition at $7.124
  \times 10^{-6}$ but for higher energies, an ever-increasing
  neighbourhood of initial conditions surrounding $\bu_s$ exists which
  trigger transition (indicated by the jump in $G$).
}
\label{fig:bubble}
\end{figure}

\begin{figure}
\centering
\resizebox{0.9\textwidth}{!}{\includegraphics{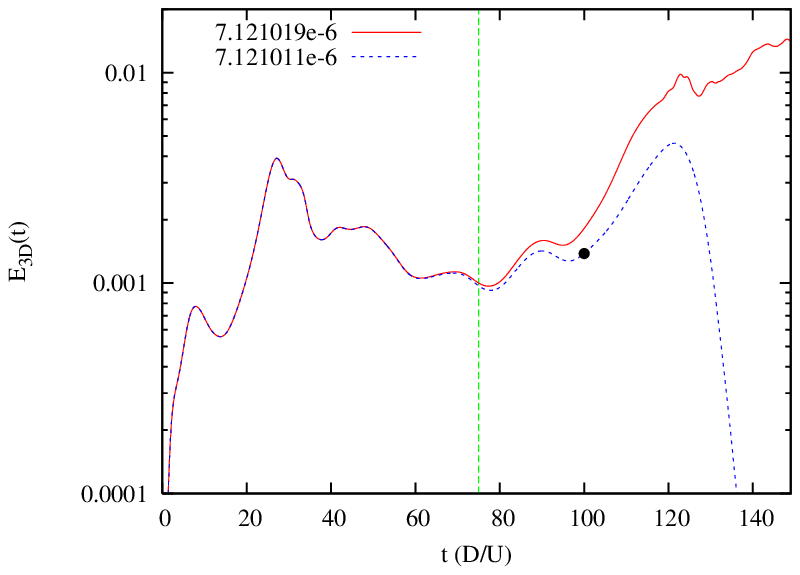}}
\caption{Trajectories close to the laminar-turbulent boundary, or
  `edge'. Nearby initial conditions are $\vec{u}_{ic}=A\vec{u}_s$,
  with $A$ selected to give the indicated energies. The case
  $E_0=7.121011 \times 10^{-6}$ relaminarises but not before passing
  close by the edge attractor (marked by the dot which is the energy
  of the $S1$ travelling wave embedded in it). The case $E_0=7.121019
  \times 10^{-6}$ tracks the edge before leading to turbulence, The
  vertical line is $T=T_{opt}=75D/U$. }
\label{fig:edge}
\end{figure}

In order to assess Conjecture 2, we now consider the behaviour of the
optimal solution close to the edge. Figures \ref{fig:phaseConv1} and
\ref{fig:phaseConv2} already provide some evidence that the NLOP at
$7.058\times 10^{-6}$ and the turbulent seed found at $7.124 \times
10^{-6}$ are similar at least in terms of their axisymmetric and
streamwise-independent energy fractions. In order to probe the
accuracy of Conjecture 2 further, we look at the one initial condition
that in section \ref{sec:longTime} our algorithm identified for
$E_0=7.124\times10^{-6}$ that was a turbulent seed, $\bu_s$
(indicated by the circle in figure \ref{fig:growth2400}) .  The fact
that only the one condition was found above the edge of chaos suggests
that in this region there is only a very small set of
turbulence-triggering initial conditions. We attempt to quantify this
by considering the evolution of initial conditions of the form
\begin{equation}
\bu_{ic}(A,d):=A[(1+d)\bu_{l=0}+(1-d)\bu_{l\ne 0}]
\end{equation}
%
%
where $\bu_{l=0}$ and $\bu_{l\ne 0}$ are the streamwise independent and
dependent parts of $\bu_s$ and $A$ is adjusted to give the required
value of $E_0$. The amount of growth after $75D/U$ is shown as a
function of $E_0$ and $d$ in figure \ref{fig:bubble}. The jump in
growth from $O(10^3)$ to $O(10^4)$ clearly demarcates where the edge
of chaos is crossed. The narrowness of the peak for $E_0=7.124\times
10^{-6}$ and the observation that a mere $\sim 0.3\%$ reduction in
amplitude is enough to dip beneath the edge, indicates that $\bu_s$ is
very close to a local minimum of the edge. The precise energy at which
$\bu_{ic}(A,d=0)$ crosses the edge is plotted in figure
\ref{fig:edge}. Here two bracketing cases are shown: $E_0=7.121011
\times 10^{-6}$, which ultimately relaminarises, and $E_0=7.121019
\times 10^{-6}$, which leads to turbulence. The closeness of these
energies means that both evolutions track the edge to $T \approx
T_{opt}$ before going their separate ways. This emphasizes that to
improve the estimate of $E_c$ discussed above, $T_{opt}$ has to be
increased.  

\begin{figure}
\centering
\resizebox{0.8\textwidth}{!}{\includegraphics{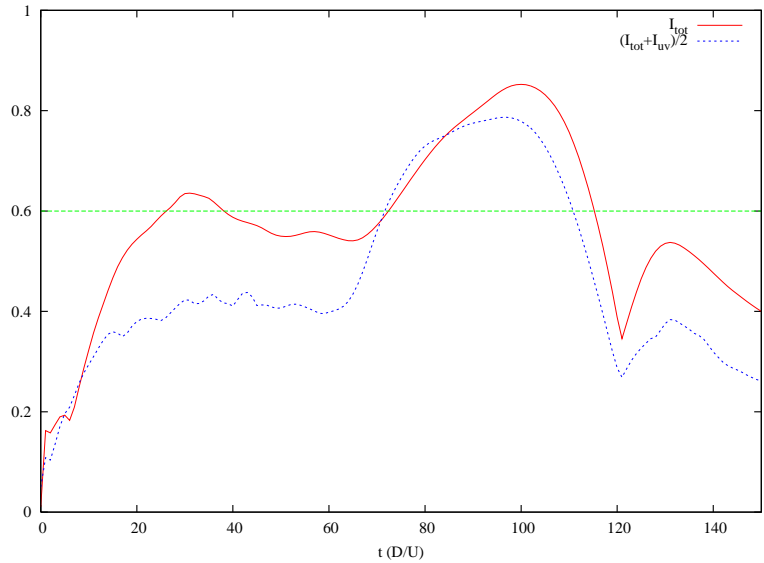}}
\caption{Correlation function data which measures how close the
  instantaneous velocity field is to the asymmetric travelling wave
  $S1$ for the trajectory which relaminarises in figure \ref{fig:edge}
  with $E_0=7.121011 \times 10^{-6}$. The fact that $I_{tot}$ and
  $\half(I_{tot}+I_{uv})$ exceed $0.75$ at $t=100 \, D/U$ indicates a
  very close visit (Kerswell \& Tutty 2007). The importance of this
  visit is that $S1$ is believed embedded in the chaotic edge state
  for the pipe length and $Re$.  }
\label{fig:approach}
\end{figure}

Coincidentally, while following the relaminarising case, the flow was
found to transiently resemble the asymmetric travelling wave believed
to be embedded in the edge state (Pringle \& Kerswell 2007, later
named $S1$ in Pringle et al. 2009) at $t \approx 100 \, D/U$. This was
verified by calculating the two correlation functions, $I_{tot}$ and
$\half (I_{tot}+I_{uv})$, introduced in Kerswell \& Tutty (2007,
definitions (2.3) and (2.5)). Figure \ref{fig:approach} shows that
both these correlations simultaneously exceed 0.75 at $t \approx 100
\, D/U$ clearly indicating a very close `visit' (0.6 was deemed good
enough to indicate a `close' visit by Kerswell \& Tutty 2007).  The
fact that this visit takes place is not a surprise but more a check of
consistency: the edge state is believed unique and therefore a global
attractor on the edge at this $Re$ and pipe length (Schneider,
Eckhardt \& Yorke 2007). It is worthy of note, however, that it takes a
comparatively long time of $\approx 100 \, D/U$ for a flow trajectory
starting at the lowest energy point on the edge to reach the $S1$ 
state.

\begin{figure}
\begin{center} \setlength{\unitlength}{1cm} \begin{picture}(13,4)
\put(0, 0){\epsfig{figure=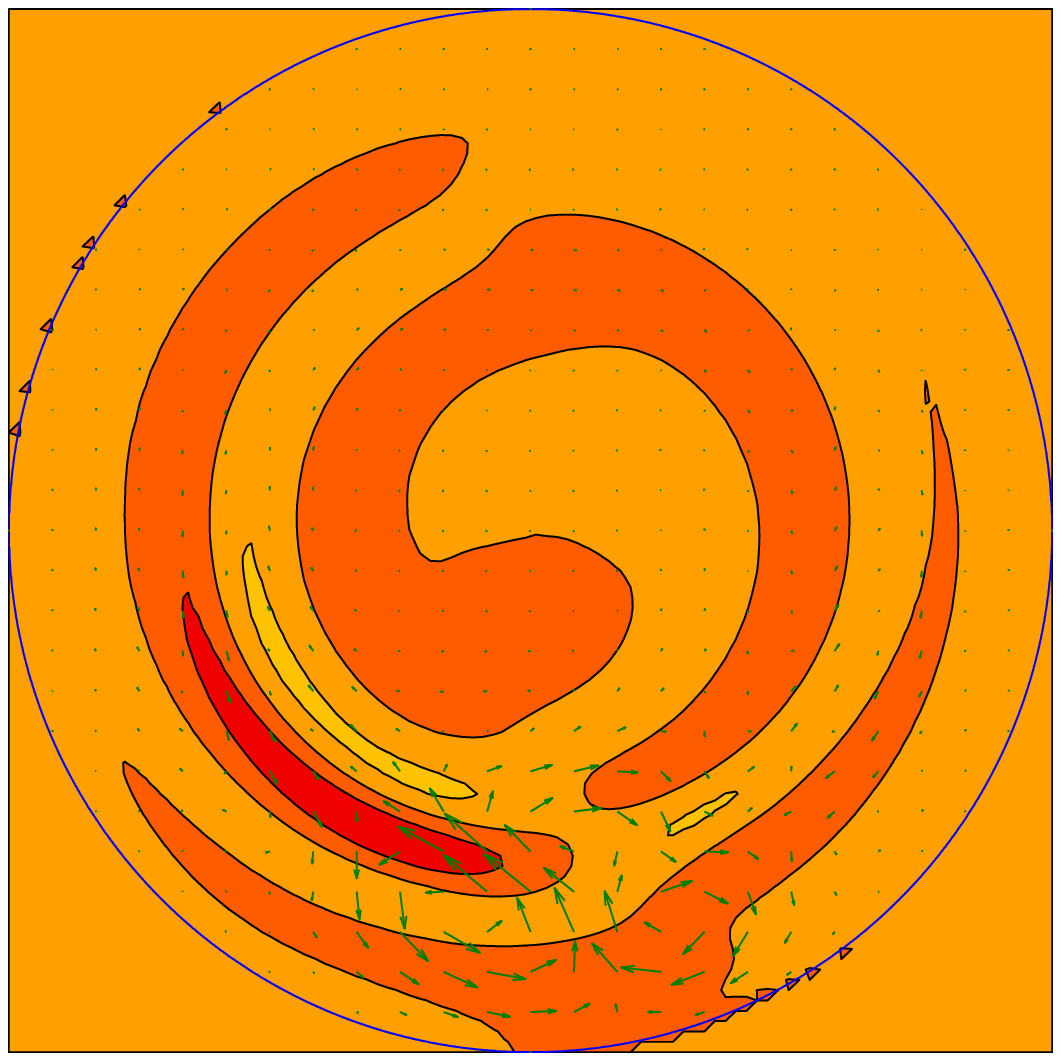,width=4cm,clip=true}}
\put(4.5, 0){\epsfig{figure=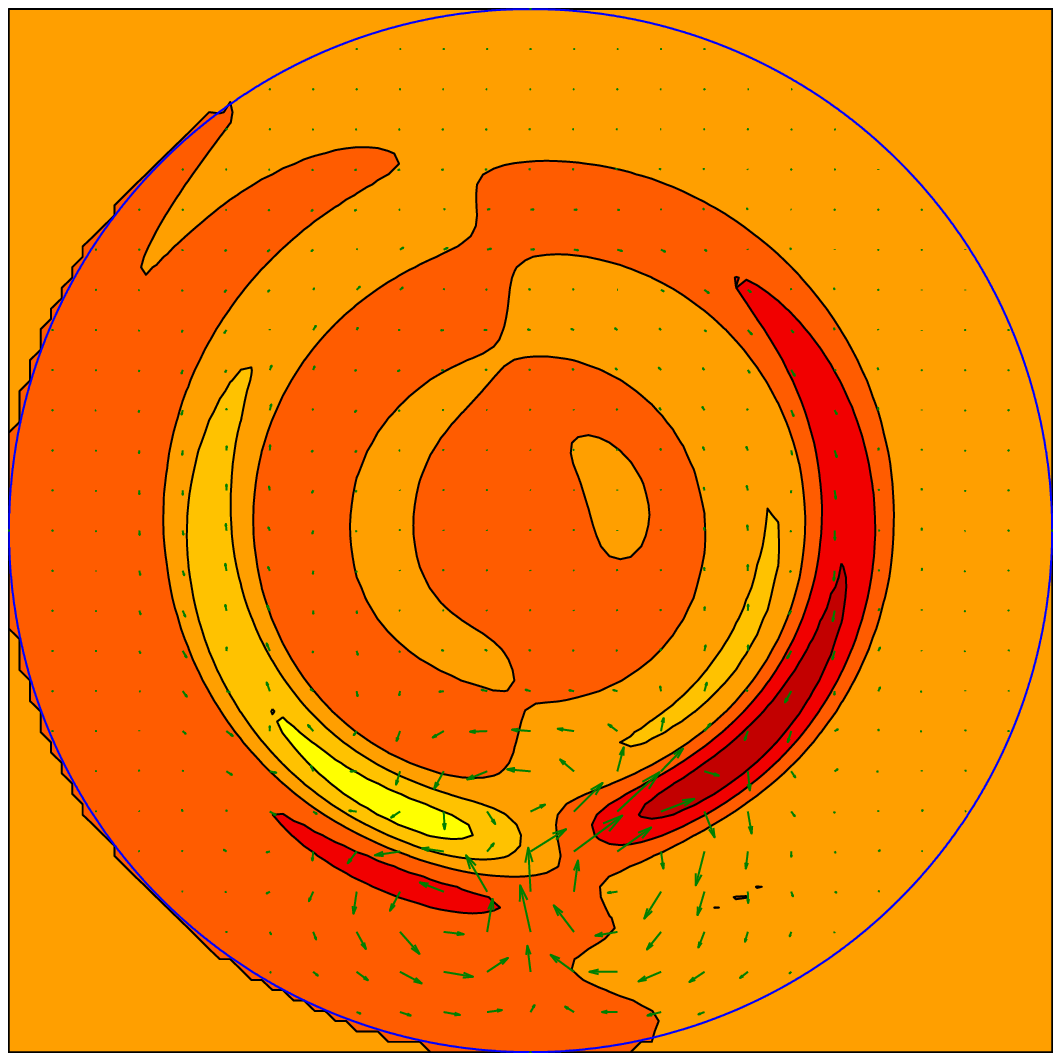,width=4cm,clip=true}}
\put(9, 0){\epsfig{figure=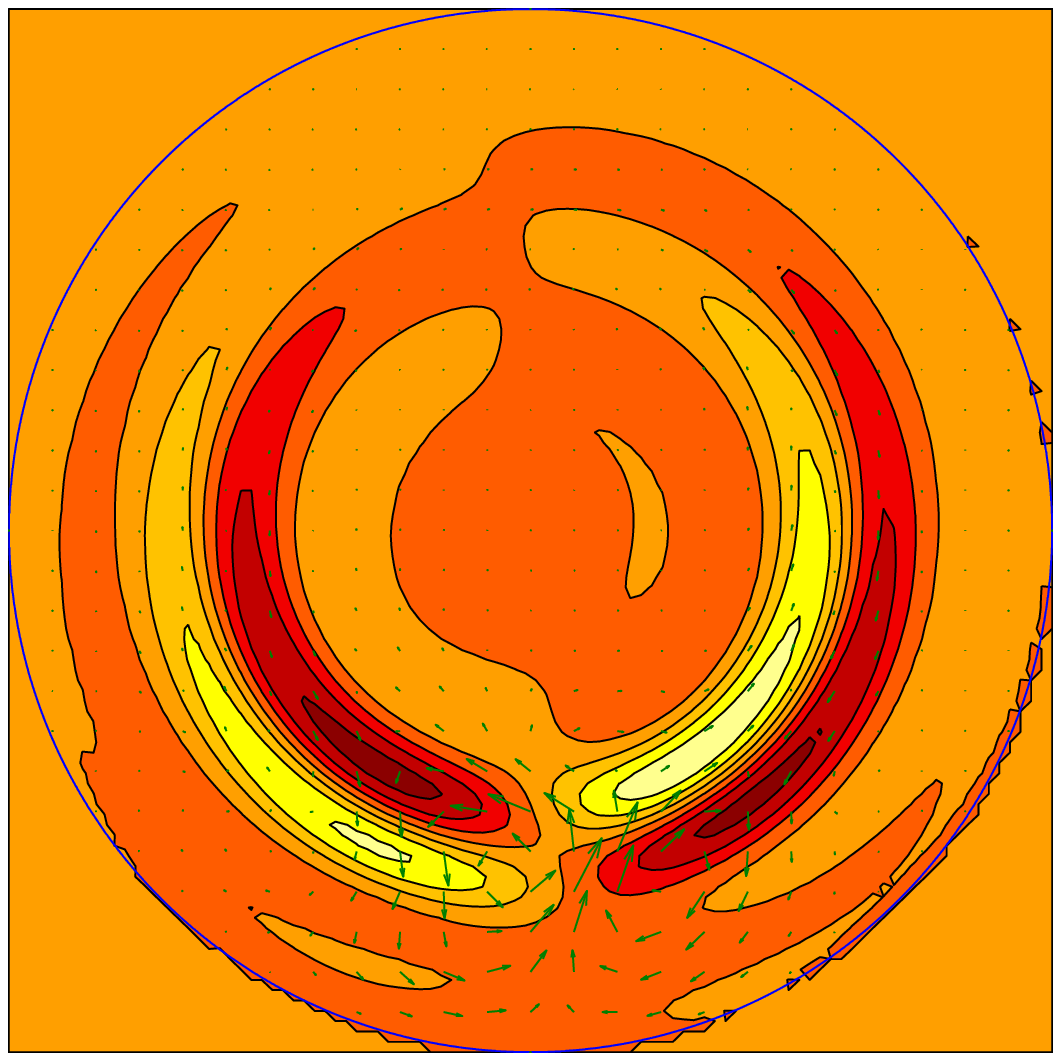,width=4cm,clip=true}}
 \end{picture} \end{center}
\caption{The nonlinear optimal calculated in section
  \ref{sec:longTime} for $E_0=7.058\times 10^{-6}$ (left) and the
  turbulent seed $\bu_s$ (right). The close similarity between the two
  solutions is striking with the turbulent seed having stronger
  streaks but an otherwise comparable structure. For comparison 
  we have also included the final state found for $E_0=7.077\times 10^{-6}$ 
  (middle). This state appears to be an intermediary between the two other 
  states.}
\label{fig:minSeed}
\end{figure}

Examining the NLOP for $E_0=7.058 \times 10^{-6}$ and $\bu_s$, it
seems reasonable to conclude that the minimal seed of turbulence is
`sandwiched' in between. The form of the two solutions are shown in
figure \ref{fig:minSeed} (along with the last iterate calculated at
$7.077 \times 10^{-6}$) and it is clear that they don't alter much as
the edge is crossed.  This supports Conjecture 2.

%
%

%
%


%
%

\section{Discussion\label{sec:conc}}
%
%
We first summarise what has been done in this paper. An exploratory
nonlinear energy growth calculation in PK10 showed that the form of
the optimal initial disturbance changes suddenly at a small
(pre-threshold) but finite initial energy level $E_{3d}$ from a global
linear optimal (weakly modified by nonlinearity) to a localised
strongly nonlinear optimal. This has been confirmed 
at higher spatial and temporal
resolution. The physical processes responsible for the enhanced energy
growth of the new nonlinear optimal (NLOP) have been identified as
three known linear growth mechanisms --- the Orr mechanism, oblique wave
transient growth and the lift-up effect --- acting sequentially and
coupled together via the nonlinearity of the Navier-Stokes equations.
These mechanisms operate on differing timescales yet appear able to
pass on their growth to the next (slower) process so that the total
growth outweighs any of their individual contributions.  The NLOP is
also localised, which is an inherently nonlinear feature designed to
cheat the global initial energy constraint. The calculations here have
also managed to converge this nonlinear optimal state beyond the
threshold energy level at which turbulence could be triggered. This
highlighted two issues for the optimisation strategy to identify this
energy threshold: 1) the optimisation time needs to be large enough
that turbulence-triggering initial conditions have time to reach the
turbulent state; and 2) the energy levels of the turbulent state need
to be above those for laminar flows so that the optimisation algorithm will
naturally seek them out.  This indicates that the optimisation strategy
discussed here is better suited to larger flow domains and more
supercritical (higher $Re$) regimes. Ironically, it is now clear that
doing exploratory (cheap) calculations in a small domain at low $Re$
was a natural but bad choice in PK10.

Calculations in a longer pipe 5D at higher $Re$ (=2400) over a larger
time period found the same general situation as in the $\half \pi D$
pipe of PK10. At low initial energies, the optimal is the (global)
linear optimal weakly modified by nonlinearity. At a certain small but
finite initial energy $E_{3d}$, a new localised 3D optimal (NLOP) is
preferred, which stays the optimal until the algorithm fails to
converge at $E_{fail}$. The only significant difference is that in the
longer $5D$ pipe, the NLOP is starting to streamwise-localise in
contrast to the $\half \pi D$ NLOP, where the shortness of the domain
prevents this. Above $E_{fail}$, initial conditions
that trigger turbulence are found to exist on the energy hypersurface
and to the accuracy available, $E_{fail}=E_c$. This supports
Conjecture 1, which presupposes that the optimisation algorithm will
find any turbulent-triggering states if they exist on the energy
hypersurface and then fail to converge as a result.  As way of
confirming this, the algorithm was tested with a variety of very
different starting conditions with the same optimal emerging,
indicating that the optimisation algorithm {\em is} able to explore
the energy hypersurface.

\begin{figure}
\centering
\resizebox{0.8\textwidth}{!}{\includegraphics{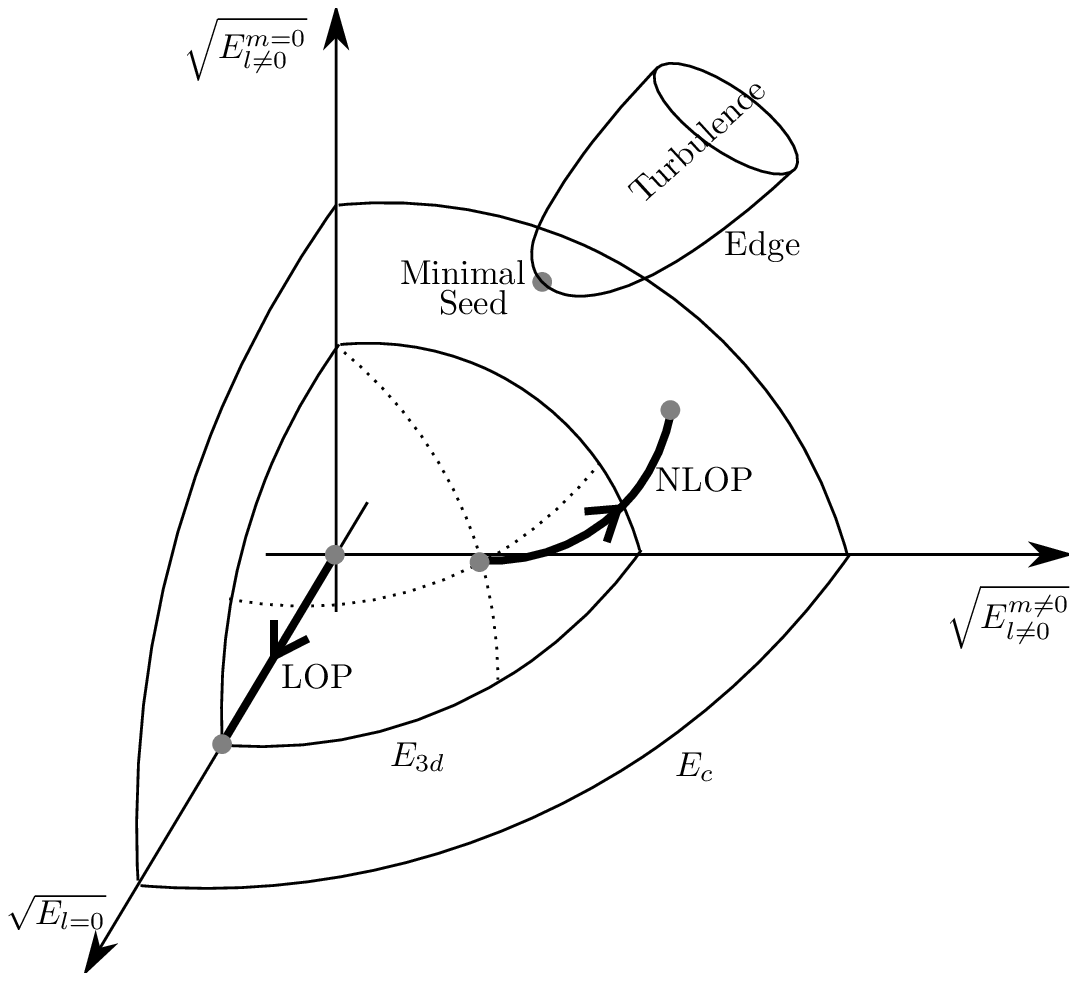}}
\caption{Cartoon to illustrate Conjectures 1 and 2. Conjecture 1
  asserts that the optimisation approach will yield a well-defined
  nonlinear optimal (NLOP) up until $E_0=E_c$ whereupon turbulence can
  be triggered by the minimal seed. Conjecture 2 asserts that the NLOP
  converges as $E_0 \rightarrow E_c^{-}$ to the minimal seed (the
  converse situation where the limiting NLOP state and the minimal
  seed are different is shown for clarity). Note the LOP is
  streamwise-independent in pipe flow and hence traces the $E_{l=0}$
  axis.}
\label{fig:cartoon}
\end{figure}

Intriguingly, good evidence was also found that NLOP $\rightarrow$
minimal seed as $E_0 \rightarrow E_c^-$ in support of the stronger
Conjecture 2 at least for this flow, geometry and $Re$. Pictorially,
this means that the NLOP for $E_c$ and the minimal seed actually
coincide in figure \ref{fig:cartoon} rather than the more general
situation shown where the two differ (for clarity). It was also argued
that, with enough computational power, the threshold energy $E_c$ and
the minimal seed could be calculated to arbitrary accuracy by
increasing the spatial and temporal resolution as well as $T_{opt}$,
which improves the algorithm's ability to discern between trajectories
that become turbulent and those that relaminarise.

%
%

%
%
The strategy advocated here for determining the minimal finite
amplitude disturbance to trigger transition to turbulence in shear
flows involves constructing and iteratively solving a variational
problem. The objective functional must be selected such that it
identifies turbulent velocity fields by taking enhanced values
compared to those for laminar fields. This is then maximised via
searching over all incompressible disturbances of fixed amplitude that
respect the boundary conditions over an asymptotically long time
period constrained by the {\em full} Navier-Stokes equations.  All of
the results discussed here have been obtained using the perturbation
energy growth over a given period as the key functional. This
certainly takes enhanced values for turbulent velocity fields in the
5D pipe at $Re=2400$ and Conjecture 1 seems to hold true. However,
other choices should also work equally well, e.g.\ the total
dissipation (Monokrousos et al. 2011), provided they share this
crucial property. If Conjecture 1 is indeed true, then the
optimisation strategy discussed here will identify the threshold
energy level $E_c$. However, more is available too, albeit indirectly,
as the minimal seed should be the unique initial condition that
triggers turbulence as $E_0 \rightarrow E_c^+$.  Given this, the
status of Conjecture 2, although conceptually fascinating, seems less
important practically. Whether or not Conjecture 2 holds for energy
growth (and, admittedly, we only have one supportive data analysis here),
it can of course be restated for any functional. Then the question
really is: is there a universal functional that when optimised always
identifies the minimal seed as $E \rightarrow E_c^-$ for a class of
flows (e.g. wall-bounded shear flows)? This seems unlikely to be
exactly true but nevertheless may be {\em approximately} true for some
subset of functionals. 
Then any of these could give acceptable
predictions depending on how the results are to be subsequently used
(e.g. designing disturbances in the laboratory). Certainly this would
seem to be the case using the energy growth functional 
given the comparison in figure \ref{fig:minSeed}.

The variational approach espoused here is, of course, incredibly
flexible. Changing the key functional is straightforward as is the
initial (norm) constraint on the competitor initial fields. Although
the discussion above has concentrated on the initial perturbation
energy $E_0$, it should be clear other norm choices can be made.
Providing the functional under consideration jumps to large values for
turbulent flows, the optimisation algorithm should converge up until
the first point (as the norm hypersurface `expands' away from the
laminar state) at which the edge penetrates the hypersurface (as per
Conjecture 1). Furthermore, the turbulent state does not have to be
the only target of the approach. Identifying the peak instantaneous
pressure in a transitional flow is a key concern for pipeline
structural integrity. One could easily imagine formulating an
optimisation problem to maximise the pressure after time $T_{opt}$
over all disturbances of initial energy $E_0$ where $T_{opt}$ is also
part of the optimisation procedure.

%
%

The long term objective of this theoretical work is to design better
(lower energy) ways to trigger turbulence with a view to informing
{\em control} techniques. Further calculations clearly need to be
carried out in more realistic geometries to see, for example, if
universal localised minimal seeds emerge. Even now though, this work
is in a position to stimulate new experiments. The NLOP identified
here indicates that structures that initially point into the shear
will outgrow the equivalent structure directed across the shear (the
Orr mechanism). This suggests a modification of the recent experiments
of Peixinho \& Mullin (2007) which were designed to generate oblique
rolls by blowing and sucking directly across the shear. The
calculations performed here indicate that their threshold scaling
exponent (a non-trivial flux $\sim Re^{-1.5}$) for transition may
possibly be further reduced if the blowing and sucking is inclined
upstream to take advantage of the Orr mechanism.

\section{Glossary\label{sec:glossary}}

\begin{tabular}{l p{10cm}}
$E_0$ & initial energy of a perturbation \\[0.1cm] 
$E_T$ & final energy of a perturbation after time T \\[0.1cm] 
$E_{3d}$ & initial energy at which the NLOP first emerges as 
the new optimal\\[0.1cm] 
$E_{fail}$ & the
  minimum energy for which the optimisation routine fails to converge
\\[0.1cm] 
$E_c$ & the critical energy corresponding the minimum
  energy of the edge \\[0.1cm] 
$\mathscr{E}_c(\mathbf{u})$ & the
  minimum energy of a perturbation of the form $A\mathbf{u}$ required
  to trigger turbulence \\[0.1cm] 
$T_{opt}$ & the target time in an
  optimisation procedure \\[0.1cm] 
$T_{lin}$ & the time for which
  transient growth is maaximised in the linear problem \\[0.1cm]
$T_{turb}$ & typical time period required for the onset of
  turbulence \\[0.1cm] 
$\mathbf{u}_{2D}(\mathbf{x};Re,E,L,T)$ & the
  two dimensional optimal for the Reynolds number, energy, domain
  length and optimisation time T specified \\[0.1cm]
$\mathbf{u}_{3D}(\mathbf{x};Re,E,L,T)$ & the three dimensional
  optimal for the Reynolds number, energy, domain length and
  optimisation time T specified
\end{tabular}

\begin{acknowledgements}
The authors acknowledge insightful conversations with Brian Farrell,
Petros Ioannou and Dan Henningson. The calculations in this paper
were carried out at the Advanced Computing Research Centre, University
of Bristol.
\end{acknowledgements}


\vspace{1cm}

\begin{thebibliography}{}
%
%
\bibitem[Andersson, Berggren \& Henningson (1999)]{Andersson}
  \textsc{Andersson, P., Berggren, M. \& Henningson, D. S. } 1999
  Optimal disturbances and bypass transition in boundary layers
  \emph{Phys. Fluids} \textbf{11}, 134--150.
%

%
%
\bibitem[Butler \& Farrell (1992)]{Butler} \textsc{Butler, K. M. \& Farrell,
  B. F.}  1992 3-Dimensional optimal perturbations in viscous shear-flow
  \emph{Phys. Fluids} \textbf{4}, 1637--1650.
%
%
\bibitem[Cherubini et al]{Cherubini10} \textsc{Cherubini,
  S., De Palma, P., Robinet, J.-Ch. \& Bottaro, A.}  2010 
Rapid path to transition via nonlinear localized optimal perturbations
in a boundary-layer flow. \emph{Phys. Rev. E} \textbf{82}, 066302.
%
%
\bibitem[Corbett and Bottaro (2000)]{Bottaro} \textsc{Corbett, P. \& 
Bottaro, A.}  2000 Optimal perturbations for boundary layers subject
  to streamwise pressure gradient.  
\emph{Phys. Fluids} \textbf{12}, 120--130.
%
\bibitem[Duguet, Willis \& Kerswell (2008)]{Duguet08} \textsc{Duguet,
  Y., Willis, A.P. \& Kerswell, R.R.}  2008 Transition in pipe flow:
  the saddle structure on the boundary of turbulence.  \emph{J. Fluid
    Mech.} \textbf{613}, 255-274.
%
\bibitem[Duguet, Brandt \& Larsson (2010)]{Duguet10} \textsc{Duguet,
  Y., Brandt, L. \& Larsson, B.R.J.}  2010 Towards minimal
  perturbations in transitional plane Couette flow. \emph{Phys. Rev. E} 
\textbf{82}, 026316.
%
\bibitem[Eggels et al]{Eggers} \textsc{Eggels, J.G.M., Unger, F.,
  Weiss, M.H., Westerweel, J., Adrian, R.J., Friedrich, R. \&
  Nieuwstadt, F.T.M.}  1994 
Fully-developed turbulence pipeflow - a comparison between direct
numerical simulation and experiment
  \emph{J. Fluid Mech.} \textbf{268}, 175--209.
%
\bibitem[Farrell \& Ioannou]{Farrell} \textsc{Farrell, B.F. \&
  Ioannou, P.J.}  1993 Optimal excitation of three-dimensional
  perturbations in viscous constant shear flow \emph{Phys. Fluids}
  \textbf{5}, 1390--1400.
%
\bibitem[Gustavsson (1991)]{Gustavsson} \textsc{Gustavsson, L. H.}
  1991 Energy growth of 3-dimensional disturbances in plane Poiseuille
  flow \emph{J. Fluid Mech.} \textbf{224}, 241--260.
%
\bibitem[Itano \& Toh]{Itano} \textsc{Itano, T. \& Toh, S.}  2001 The
  dynamics of bursting process in wall turbulence
  \emph{J. Phys. Soc. Japan.} \textbf{70}, 703--716.
%
\bibitem[Kerswell \& Tutty]{KT07} \textsc{Kerswell, R.R. \& Tutty,
  O.R.}  2007 
Recurence of travelling waves in transitional pipe flow
  \emph{J. Fluid Mech.} \textbf{584}, 69--102.
%
\bibitem[Luchini \& Bottaro]{Luchini} \textsc{Luchini, P. \& Bottaro, A.}  1998 
  G\"ortler vortices: a backward-in-time approach to the receptivity problem
  \emph{J. Fluid Mech.} \textbf{363}, 1--23.

\bibitem[Luchini]{Luchini2000} \textsc{Luchini, P.}  2000 
Reynolds-number-independent instability of the boundary layer over a
flat surface: optimal perturbations
  \emph{J. Fluid Mech.} \textbf{404}, 289--309.
%
\bibitem[Meseguer \& Trefethen]{Meseguer} 
\textsc{Meseguer, A. \& Trefethen, L.N.}  2003 
Linearized pipe flow to Reynolds number $10^7$
  \emph{J. Comp. Physics} \textbf{186}, 178--197.
%
\bibitem[Monokrousos et al.]{Monokrousos} \textsc{Monokrousos, A.,
  Bottaro, A., Brandt, L., Di Vita, A.  \& Henningson, D.S.}  2011
  Nonequilibrium thermodynamics and the optimal path to turbulence in
  shear flows \emph{Phys. Rev. Lett.} \textbf{106}, 134502.
%
\bibitem[Orr (1907)]{Orr} \textsc{Orr, W.M.F.}  1907
The stability or instability of the steady motions of a perfect liquid
and of a viscous liquid. Part I: A perfect liquid. Part II: A viscous liquid.  
  \emph{Proc. R. Irish Acad. A} \textbf{27}, 9--138.
%
\bibitem[Peixinho \& Mullin (2007)]{PM07} \textsc{Peixinho,
  J. \& Mullin, T.}  2007 Finite-amplitude thresholds for transition
  in pipe flow
  \emph{J. Fluid Mech.} \textbf{582}, 169-178.
%
\bibitem[Pringle \& Kerswell (2007)]{PK07} \textsc{Pringle,
  C.C.T. \& Kerswell, R.R.}  2007 Asymmetric, helical and mirror-symmetric
  travelling waves in pipe flow
  \emph{Phys. Rev. Lett.} \textbf{99}, 074502 (referred to as PK07 in
  the text).
%
\bibitem[Pringle, Duguet \& Kerswell (2009)]{PDK09} \textsc{Pringle,
  C.C.T., Duguet, Y.  \& Kerswell, R.R.}  2009 
Highly symmetric travelling waves in pipe flow  
\emph{Phil. Trans. Roy. Soc. A.} \textbf{367}, 457-472.
%
\bibitem[Pringle \& Kerswell (2010)]{Pringle} \textsc{Pringle,
  C.C.T. \& Kerswell, R.R.}  2010 Using nonlinear transient growth to
  construct the minimal seed for shear flow turbulence.
  \emph{Phys. Rev. Lett.} \textbf{105}, 154502 (referred to as PK10 in
  the text).
%
\bibitem[Reddy et al. (1998)]{Reddy} \textsc{Reddy, S.C., Schmid,
  P.J., Baggett, J.S. \& Henningson, D.S.}  1998 On the stability of
  streamwise streaks and transition thresholds in plane channel flows.
  \emph{J. Fluid Mech.} \textbf{365}, 269--303.
%
\bibitem[Reddy and Henningson (1993)]{Reddy&Henningson} \textsc{Reddy,
  S.C. \& Henningson, D.S.}  1993 Energy growth in visocus channel
  flows \emph{J. Fluid Mech.} \textbf{252}, 209--238.
%
\bibitem[Schmid \& Henningson (1994)]{Schmid 1994}
  \textsc{Schmid, P.J. \& Henningson, D.S.} 1994
 Optimal energy growth in Hagen-Poiseuille flow
 \emph{J. Fluid Mech.} \textbf{277}, 197.
%
\bibitem[Schneider, Eckhardt \& Yorke (2007)]{Schneider07}
  \textsc{Schneider, T.M., Eckhardt, B. \& Yorke, J.A.}  2007
 Turbulence transition and the edge of chaos in pipe flow
 \emph{Phys. Rev. Lett.} \textbf{99}, 034502.
%
\bibitem[Schneider \& Eckhardt]{Schneider09}
  \textsc{Schneider, T.M. \&  Eckhardt, B.}  2009
 Edge states intermediate between laminar and turbulent dynamics in pipe flow
 \emph{Phil. Trans. R. Soc. A} \textbf{367}, 577-587.
%
\bibitem[Skufca, Yorke \& Eckhardt (2006)]{Skufca} \textsc{Skufca,
  J.D., Yorke, J.A. \& Eckhardt, B.}  2006 Edge of Chaos in a Parallel
  Shear Flow.  \emph{Phys. Rev. Lett.} \textbf{96}, 174101.
%
\bibitem[Trefethen et al. (1993)]{Trefethen} \textsc{Trefethen, L. N.,
  Trefethen, A. E. \& Reddy, S. C.}  1993
Hydrodynamic stability without eigenvalues
  \emph{Science} \textbf{261}, 578--584.
%
\bibitem[Viswanath \& Cvitanovic (2009)]{Viswanath} \textsc{Viswanath,
  D.. \& Cvitanovic, P.}  2009 Stable manifolds and the transition to
  turbulence in pipe flow
  \emph{J. Fluid Mech.} \textbf{617}, 215-233.
%
\bibitem[Willis & Kerswell(2009)]{Willis} \textsc{Willis, A.P. \&
  Kerswell, R.R.}  2009 Turbulent dynamics of pipe flow captured in a
  reduced model: puff relaminarisation and localised `edge' states.
  \emph{J. Fluid Mech.} \textbf{619}, 213-233.
%
\bibitem[Zuccher et al. (2004)]{Zuccher} \textsc{Zuccher, S., 
  Luchini, P. \& Bottaro, A.}  2004 Algebraic growth in a Blasius boundary layer: 
  optimal and robust control by mean suction in the nonlinear regime.   \emph{J. Fluid Mech.} 
  \textbf{513}, 135--17.
%

\end{thebibliography}
\end{document}